\documentclass[sigconf]{acmart}

\usepackage{booktabs} 
\usepackage{color}
\usepackage[utf8]{inputenc}
\usepackage{url}
\usepackage{multirow}
\usepackage{siunitx}
\usepackage{array}
\usepackage{subcaption}
\usepackage[moderate]{savetrees}
\usepackage{float}

\newcolumntype{R}{>{\raggedleft\arraybackslash}p{1.5cm}}

\copyrightyear{2018}
\acmYear{2018}
\setcopyright{acmcopyright}
\acmConference[DH'18]{2018 International Digital Health Conference}{April 23--26, 2018}{Lyon, France}
\acmBooktitle{DH'18: 2018 International Digital Health Conference, April 23--26, 2018, Lyon, France}
\acmPrice{15.00}
\acmDOI{10.1145/3194658.3194667}
\acmISBN{978-1-4503-6493-5/18/04}

\begin{document}
\title{Predicting antimicrobial drug consumption using web search data}

\author{Niels Dalum Hansen}
\authornote{Work done while at IBM Denmark and University of Copenhagen, Denmark}
\affiliation{%
  \institution{Statens Serum Institut, Denmark}
}
\email{nidh@ssi.dk}
\author{Kåre Mølbak}
\affiliation{%
	\institution{Statens Serum Institut, Denmark}
}
\email{krm@ssi.dk}

\author{Ingemar J. Cox}
\authornote{Also with Department of Computer Science, University College London, UK}
\affiliation{%
	\institution{University of Copenhagen, Denmark}
}
\email{ingemar.cox@di.ku.dk}
\author{Christina Lioma}
\affiliation{%
	\institution{University of Copenhagen, Denmark}
}
\email{c.lioma@di.ku.dk}

\begin{abstract}

Consumption of antimicrobial drugs, such as antibiotics, is linked with antimicrobial resistance. Surveillance of antimicrobial drug consumption is therefore an important element in dealing with antimicrobial resistance. Many countries lack sufficient surveillance systems. Usage of web mined data therefore has the potential to improve current surveillance methods. To this end, we study how well antimicrobial drug consumption can be predicted based on web search queries, compared to historical purchase data of antimicrobial drugs. We present two prediction models (linear Elastic Net, and non-linear Gaussian Processes), which we train and evaluate on almost 6 years of weekly antimicrobial drug consumption data from Denmark and web search data from Google Health Trends. We present a novel method of selecting web search queries by considering diseases and drugs linked to antimicrobials, as well as professional and layman descriptions of antimicrobial drugs, all of which we mine from the open web. We find that predictions based on web search data are marginally more erroneous but overall on a par with predictions based on purchases of antimicrobial drugs. This marginal difference corresponds to $<1$\% point mean absolute error in weekly usage. Best predictions are reported when combining both web search and purchase data.

This study contributes a novel alternative solution to the real-life problem of predicting (and hence monitoring) antimicrobial drug consumption, which is particularly valuable in countries/states lacking centralised and timely surveillance systems.

\end{abstract}

%
%




\maketitle

\section{Introduction}

Surveillance of antimicrobial drug consumption, such as antibiotics, is an important element in dealing with antimicrobial resistance. Antimicrobial resistance is recognized as a major challenge, not only for the health care system, but also for economic growth and welfare \cite{oneill_review}. Use of antimicrobials is one of the main factors responsible for the development, selection and spread of antimicrobial resistance \cite{antimicrobial2017}. This has become a serious threat to public health, notably because of the emergence and spread of highly resistant bacteria, and because there are very few novel antimicrobial agents in the research and development pipeline.

In the European Union the European Surveillance of Antimicrobial Consumption Network (ESAC-Net) \cite{esac-net} is collecting reference data from national antimicrobial drug consumption surveillance systems. The availability of national and EU-wide surveillance data has been a driving factor for the political commitment necessary for successful campaigns for responsible antimicrobial drug use \cite{who-antimirocbial-drug-use}. The quality and granularity of the data varies widely even between European countries. Some countries, such as Denmark \cite{danmap}, have detailed surveillance systems that keep track of antimicrobial drug use both in primary care and hospitals with a coverage of approximately 97\% of the total usage. While others, such as Germany \cite{antimicrobial2012}, base their antimicrobial drug surveillance system on reimbursement data from insurance companies with a coverage of 85\% the usage, but only for primary care. ESAC-Net is good example of the emerging focus on surveillance of antimicrobial drug use, and the importance of surveillance data. But many counties outside of the EU, such as the US, still lack nationwide surveillance systems, and many others have no monitoring at all.  Hence, methods that can be implemented quickly and cost effectively are of great value.

Monitoring of antimicrobial drug consumption has several use cases in public health. Here we list two examples: 1) Knowing the consumption pattern of antimicrobials can be used as leverage in political discussions. Being able to document the problem and show measurable improvements can make a difference when discussing the allocation, or maintenance, of resources. 2) Identification of misuse is easier with access to detailed information about  use patterns for antimicrobial drugs. An example could be if unusual quantities of macrolides, often used for mycoplasma pneumoniae, are being prescribed in periods with low mycoplasma pneumoniae incidence. This could indicate drug misuse, i.e. people are being treated for mycoplasma pneumoniae without being infected. In such a case it could be necessary to inform doctors on correct usage of macrolides.

To improve awareness and stimulate prudent use of antimicrobial drugs, monitoring is important. We hypothesize that antimicrobial drug consumption can be predicted, and hence monitored, from online behavior, such as the queries submitted to web search engines. This can benefit public health by: (i) Allowing countries without access to real-time data to forecast time trends and seasonal patterns of consumption; (ii) allowing all countries to analyze determinants of use, e.g. which types of web search queries are important as predictors of certain classes of antimicrobial drugs. This information can be used in communication efforts to stimulate antimicrobial stewardship; (iii) complementing syndromic surveillance, e.g. web searches that are predictive of drugs for respiratory infections can be used as an indicator of these diseases.

In this paper we study how well antimicrobial drug consumption can be predicted based on web search queries, and specifically the number of submitted queries to online search engines, e.g. how frequently people have searched for ``fever'' on Google in a specific time interval. To our knowledge such web search data has not been previously used to predict antimicrobial drug consumption. However, this type of web search data has been used previously to predict other health events, e.g. influenza like illness (ILI) \cite{eysenbach2006infodemiology} or vaccination uptake \cite{www17}. We compare web search based prediction to the more traditional method of predicting based on historical purchase data of antimicrobial drugs. We present two prediction models (a linear one, namely Elastic Net, and a non-linear one, namely Gaussian Processes), which we train and evaluate on almost 6 years of weekly antimicrobial drug consumption data from Denmark and web search data mined from Google Health Trends for the location of Denmark. We further present a novel method of selecting web search queries by considering diseases and drugs linked to antimicrobials, as well as professional and layman descriptions of antimicrobial drugs, all of which we mine from the open web. We find that the prediction error of swapping historical antimicrobial drug purchase data to web search queries is overall negligible, across different prediction offsets.

%
%
%
\section{Related work}

There is a large amount of work on using web search data to predict health events, though not antimicrobial drug consumption. Focusing on work related to public health, considerable effort has been used on estimating the incidence of various diseases based on web search query frequencies. Influenza like illness (ILI) prediction has been the subject of numerous papers \cite{eysenbach2006infodemiology, polgreen2008using, ginsberg2009detecting, lazer2014parable, santillana2014can, www17_lampos, www17}, but other infectious diseases have also been predicted using web search data, e.g. dengue fever, gastrointestinal diseases, HIV/AIDS, scarlet fever, tuberculosis \cite{Bernardo}. The domain is not restricted to infectious diseases: other papers have shown that web search data can be used for prediction of vaccination uptake \cite{DalumHansen:2016:ELV:2983323.2983882, www17}, hospital admissions \cite{Agarwal} and dietary habits \cite{west2013cookies}. On an individual level, health events as diverse as pregnancy, allergy, eating disorder and post-traumatic stress disorder have also been identify based on web search query frequency analysis \cite{yom2015automatic}, illustrating the range and diversity of predictors available through web search frequency data.

We have not identified any studies on prediction of national drug consumption using web search frequency data. This does not mean that drugs have not been included in the previous studies. Inspecting the list of queries used in the prediction models for ILI reveals that brand names for cough medicine, such as Tessalon \cite{dalum2017seasonal}, Tylenol \cite{www17_lampos} or Robitussin \cite{santillana2014can}, are included as predictors of ILI. This indicates that in the case of illness people query the web for information on the relevant medication \cite{DragusinPLLJCHIW13}, leading us to the hypothesis that drug consumption can be predicted based on web search query frequencies.

Prediction with web search query frequency data can be divided into two steps: (i) query selection and (ii) prediction using query frequency data. Query selection can for example be performed using hand picked seed words \cite{eysenbach2006infodemiology, polgreen2008using}, which are used to filter relevant from irrelevant searches, or using written descriptions of the event that is being predicted \cite{DalumHansen:2016:ELV:2983323.2983882, www17}. Perhaps the most popular approach is to use  the historical correlation between the query search frequency time series and the time series to be predicted \cite{ginsberg2009detecting, lazer2014parable, santillana2014can}. When predicting using query frequency data, the most prevalent prediction models are linear models \cite{ginsberg2009detecting, lazer2014parable, santillana2014can}, but other non-linear models such as random forests \cite{www17, Agarwal} or Gaussian Processes \cite{lampos2015advances, www17_lampos} have also been used. While all these studies use web search data, other types of online data have also been used, e.g. social media data, such as Twitter messages \cite{Lampos2010a, paul2014twitter, Signorini2011}. In this work we use web search frequency data and make predictions using both a linear model and Gaussian Processes. We select queries based on a collection of written resources on antimicrobial drug consumption.

While there is, to our knowledge, no prior work on the prediction of antimicrobial drug consumption using web search data, there has been work in computational epidemiology regarding antibiotics on Twitter. In 2010 Scanfeld et al. \cite{scanfeld2010dissemination} analyzed 1000 tweets mentioning antibiotics and categorized them into 11 categories. The top three categories were: ``general use'', ``advice/information'' and ``side effects/negative reactions''. Scanfeld et al. concluded that social media was used for sharing information about antibiotics and that the tweets could be used to identify potential misuse and misunderstandings regarding antibiotics. Later, in 2014, Dyar et al. \cite{Dyar2014} made a large scale analysis of worldwide Twitter activity mentioning antibiotics in the period September 2012 to 2013. They limited their analysis to four peaks in the twitter activity and examined the reason for those peaks. They concluded that the peaks were caused by institutional events, such as public announcements from the UK Chief Medical Officer regarding antibiotics. The peaks did not result in any sustained twitter activity, and activity was generally back to baseline level after two days. Kendra et al. \cite{Kendra2015} showed in 2015 that tweets regarding antibiotic usage could be categorized automatically using a neural network. Like Dyar et al. \cite{Dyar2014}, they also observed that peaks in activity were correlated with public events such as a speech by the British prime minister and an executive order from the President of the US regarding antibiotic resistance. None of the studies address a potential relationship between Twitter activity related to antimicrobials and antimicrobial drug consumption. Since none of the studies collected data for more than one year, long term relationships between online activity and antimicrobial drug consumption have not been analyzed. In contrast, we use web search data spanning 5 years and 10 months.

Next we describe, the data collected for our analysis (Section \ref{s:data}), and then our prediction methods (Section \ref{s:prediction}).

\section{Data}
\label{s:data}

We use three categories of data: (1) Sales of antimicrobials in Denmark collected by Danish health officials; (2) Web search query frequency data from Denmark; (3) Freely available online material related to antimicrobial drugs such as disease descriptions or information about antimicrobials. We describe these next.

\subsection{Antimicrobial usage in Denmark}
\label{ss:antibioticData}

\begin{figure}
	\centering
	\includegraphics[width=8cm]{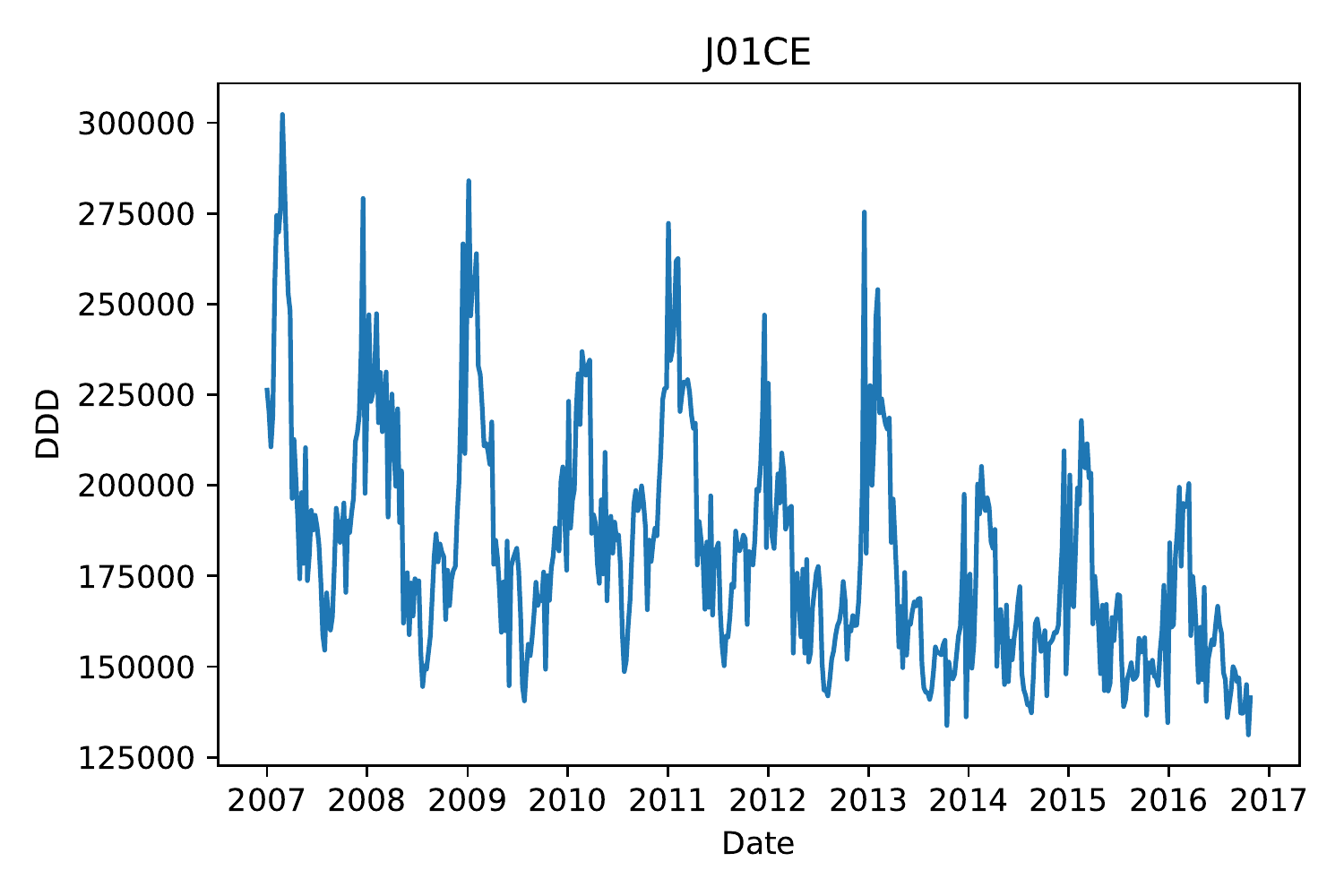}
	\caption{Weekly usage of antimicrobial J01CE in Denmark from 2007 to 2016. DDD denotes \textit{defined daily doses}.}
	\label{fig:j0ce_usage}
\end{figure}

We use weekly data on purchases of antimicrobials for people in Denmark provided by the Danish Health Data Authority from the Register of Medicinal Product Statistics. 
This covers all purchases of antimicrobials by people in Denmark, except for private hospitals and clinics (very few exist in Denmark), psychiatric hospitals, specialized non-acute care clinics, rehabilitation centers, and hospices. These exempted cases account for approximately 3\% of the total antimicrobial drug consumption.
The data spans the period 1 January 2007 – 23 October 2016, inclusive. Due to limitations on the web frequency data we only use data from 2011 onwards in the evaluation. The data is collected from pharmacies in Denmark and consists of sales data for several antimicrobial subgroups. Different subgroups are used for different diseases. We focus on the largest subgroup, namely \textit{beta-lactamase sensitive penicillins}, with the Anatomical Therapeutic Chemical (ACT) classification system code J01CE. Usage is quantified as \textit{defined daily doses} (DDD), meaning assumed average maintenance dose per day for adults for the condition the drug is registered for. Figure \ref{fig:j0ce_usage} shows the sales data of J01CE for the study period. We see that DDD tends to peak once or twice per year, and that from 2014 until 2016 the overall yearly range of DDD has decreased.   There is a noticeable change in usage from 2014 onwards. This change is probably due to national campaigns aimed at reducing usage and change in usage to other antimicrobials. J01CE is often used for treatment of pneumonia and seasonal variations are therefore expected.

\subsection{Web search query frequency}
\label{ss:queryData}
Our second category of data consists of web search queries and their frequencies. We retrieve them from the Google Health Trends API. This is an API maintained by Google and it is similar to Google Trends. Google Health Trends makes it possible to submit a query and receive aggregated weekly web search frequency data, i.e. a time series corresponding to how many times people have searched for that specific query each week. The Google Health Trends API is based on a uniform sample of 10\%-15\% of Google web searches. The results correspond to the probability of a short web search session matching the submitted query. It is possible to restrict the search both with respect to a time period and geographical region. We restrict our search to data from the period 2 January 2011 – 23 October 2016, inclusive, and for the geographical region Denmark. We only use data from 2011 and onwards because Google changed their geographical identification in 2011.

\subsection{Online antimicrobial material}
\label{sec:data_query_selection}
Our third category of data consists of freely available online information on antimicrobial drugs, and specifically on: (i) disease names, (ii) drug names, and (iii) descriptions of antimicrobials. For each of these, we extract data from the websites described below. We collect online data on antimicrobials to use them for selecting web search queries. We describe precisely how we do this in Section \ref{ss:querySelection}.

\paragraph{3.3.1. Disease names:} Descriptions of diseases are downloaded from two online resources, both maintained by sundhed.dk (ENG: health.dk) a governmental web site functioning as a digital gateway for citizens to health services, e.g. electronic patient journals, hospital treatment records, etc. The two websites, Patienthåndbogen\footnote{https://www.sundhed.dk/borger/patienthaandbogen/} (ENG: The Patient's Handbook) and Lægehåndbogen\footnote{https://www.sundhed.dk/sundhedsfaglig/laegehaandbogen/} (ENG: The Doctor's Handbook), are designed as encyclopedias of diseases. The target audience for The Patient's Handbook is laymen, and for The Doctor's Handbook health professionals.

\paragraph{3.3.2. Drug names:} The organization Dansk Lægemiddel Information A/S (ENG: Danish Drug Information) maintains two websites, min.medicin\footnote{\url{http://min.medicin.dk/}} (ENG: My Medicine), and pro.medicin\footnote{\url{http://pro.medicin.dk/}} (ENG: Pro Medicine), with descriptions of drugs available on the Danish drug market. The organization is funded by the medical industry and the Danish government. My Medicine targets laymen, while Pro Medicine targets health professionals.

\paragraph{3.3.3. Descriptions of antimicrobials:} We choose descriptions from four websites describing to laymen what antimicrobials are and when to use them: \url{www.ssi.dk}, \url{www.netdoctor.dk}, \url{www.sundhed.dk}, and \url{www.antibiotikaellerej.dk}. The four websites are maintained by the following four groups: Danish Center for Disease Control, netdoctor.dk (a leading Danish health information website), sundhed.dk, and finally a collaboration of the Danish government, the pharmacist union, the doctors union, the society for general practitioners and the Danish Center for Disease Control. We consider all of the four groups to be authoritative and neutral.

\section{Prediction of antimicrobial drug\\ consumption}
\label{s:prediction}
The goal is to predict antimicrobial drug consumption using web search data. We do this in three steps: First, we select web search queries that are likely to indicate antimicrobial drug consumption (Section \ref{ss:querySelection}); then, for each query frequency time series we generate a number of lagged versions and decide which lags should be used for the prediction (Section \ref{sec:feature_selection}); and finally we use appropriate prediction models to infer antimicrobial drug consumption (Section \ref{ss:predictionModels}).

\subsection{Query selection}
 \label{ss:querySelection}

For our analysis we use web search queries, and their frequencies, retrieved from Google Health Trends, as described in Section \ref{ss:queryData}. We select these queries based on the online antimicrobials material described in Section \ref{sec:data_query_selection} as follows.

We start with an empty set of queries and a set of seed words. Our seed words are the ATC code ``J01CE'', and the individual nouns of the antimicrobial name: ``penicillin'', ``penicilliner'' (plural form of penicillin) and ``beta-lactamase''. Using these seed words, we populate the set of queries in the following way:

\paragraph{Disease names (described in Section 3.3.1):} A disease name is added to the set of queries if the treatment description for the disease mentions one of the seed words.

\paragraph{Drugs names (described in Section 3.3.2):} A drug name is added to the set of queries if the description of the active substances mentions one of the seed words, or if the sub-heading of the drug description page contains one of the seed words.

\paragraph{Description of antimicrobials (described in Section 3.3.3):} For each of the four descriptions (found in each of the four websites described in Section 3.3.3) we extract all unique words and remove stop words, that is commonly occurring words, such as ``I'', ``and'', etc.
We use as stop word set the 100 most frequent words in CorpusDK \cite{DSL2002}, a corpus of text representing written Danish around year 2000. Based on these four sets of words, we select two partially overlapping sets of queries: (i) Words that occur in at least two descriptions of antimicrobials or antimicrobial usage targeting laymen, and (ii) words that occur in at least three descriptions of antimicrobials or antimicrobial usage targeting laymen.

The above process results in these six sets of queries:
\begin{enumerate}
	\item \textbf{Disease names pro:} Disease names used by health professionals.
	\item \textbf{Disease names lay:} Disease names used by laymen.
	\item \textbf{Drug names pro:} Drug names used by health professionals.
	\item \textbf{Drug names lay:} Drug names used by laymen.
	\item \textbf{Descriptions lay:} Words co-occurring in two descriptions of antimicrobials for laymen.
	\item \textbf{Descriptions lay frequent:} Words co-occurring in three descriptions of antimicrobials for laymen.
\end{enumerate}

\begin{table}
	\begin{tabular}{lr}
		\toprule
		Query sets & \#Queries \\
		\midrule
		Disease names pro & 47 \\
		Disease names lay & 11 \\
		Drug names pro    & 7 \\
		Drug names lay    & 8 \\
		Descriptions lay  & 72 \\
		Descriptions lay frequent & 18 \\
		\bottomrule
	\end{tabular}
	\caption{Number of queries in each query set.}
	\label{table:query_count}
\end{table}

Table \ref{table:query_count} shows the number of queries per query set. The highest number of queries is generated from \textit{Descriptions lay frequent}, followed by \textit{Disease names pro}. The queries generated from drug names (both \textit{Drug names pro} and \textit{Drug names lay}) are by far the fewest.

\subsection{Time lag selection}
\label{sec:feature_selection}
Each query in the query sets described above has an associated time series (time stamps and search frequency). It is not unlikely that there exist lagged effects, for example increased search activity in one week might correspond to increased antimicrobial drug consumption two weeks after.
To account for such effects we generate a number of lagged versions of each query frequency time series and include only a subset of them in our prediction models. We select these by fitting a linear model with the antimicrobial drug consumption data as target variable and lagged versions of the query frequencies as predictors:

\begin{equation}
	y_t = \sum_{l=1}^L \sum_{i=1}^N \beta_{(l-1) N+i} Q_{t-l, i},
\end{equation}

\noindent where $y_t$ denotes the antimicrobial drug consumption at time $t$, $L$ is the number of lags, $N$ the number of queries, $Q_{t,i}$ is the query frequency at time $t$ for query $i$, and $\beta$ is the model coefficient. Each model coefficient is related to a variable, i.e. a query and a time offset, and the size of the coefficient defines the importance of the variable in the prediction model. We use the size of the coefficients for selecting queries and corresponding lags, which is described below.

To fit the model, we use Elastic Net which combines L1 and L2 regularization. Two hyper-parameters, $\lambda_1$ and $\lambda_2$, control the L1 and L2 penalization for the Elastic Net regularization. Using matrix notation, the function being minimized can be written as:

\begin{equation}\label{eq:elasticnet}
\| y- \beta X \|^2 + \lambda_1 \| \beta \|_1 + \lambda_2 \| \beta \|^2_2
\end{equation}

\noindent where $y$ is a vector with the target values, and $X$ is a matrix with lagged query frequency time series. Elastic Net is well suited for problems where the number of variables is much larger than the number of training samples \cite{zou2005regularization}, which can be the case in our setups if many lags are used. In addition, Elastic Net groups correlated features and, either keeps them in the model, i.e. non-zero coefficient, or leaves them out \cite{zou2005regularization}. This is a useful attribute for query selection, since we would like to select all correlated queries. To select queries, we sort the queries according to the absolute value of the coefficients, and pick the 100 with highest absolute value. While Elastic Net can be used for query selection without any threshold, i.e. by removing features with zero valued coefficients, there is no upper bound on the number of features. Due to problems with model fitting of the Gaussian processes, we enforce a hard threshold of 100 queries.


\subsection{Prediction Models}
\label{ss:predictionModels}
Our goal is to predict antimicrobial drug consumption, i.e. the number of Defined Daily Doses (DDDs) of antimicrobials being consumed per week. To this end, we use two types of prediction models: (i) Linear models with Elastic Net regularization presented in Section \ref{sec:linear}. This is a common approach often used when predicting with web search data \cite{ginsberg2009detecting, yang2015accurate}. (ii) Gaussian Processes, which are capable of capturing non-linearities in the data presented in Section \ref{sec:gp}. These models have successfully been applied to web search data to improve predictive performance \cite{lampos2015advances, www17_lampos}.

For both prediction models we use three setups for our predictions:
(i) Using only web search data,
(ii) using only historical antimicrobial drug consumption data, and
(iii) combining historical antimicrobial drug consumption data and web search data.

We explain our prediction models next.

\subsubsection{Linear models for antimicrobial drug consumption prediction}
\label{sec:linear}

We use linear models because they are easy to fit and to interpret, allowing us to draw direct inferences between their output and the real life prediction problem at hand. Using only web search data, the prediction model is defined as:

\begin{equation}
	y_{t+p} = \beta_0 + \sum_{i=1}^N \beta_i Q_{t,i}
\end{equation}

\noindent where $y_{t+p}$ is the antimicrobial drug consumption data at time $t$ with a prediction offset of $p$, $N$ is the number of queries, $Q_{t,i}$ is the query frequency at time $t$ for query $i$, and the $\beta$s are the model coefficients.

When using only antimicrobial drug consumption data, we use a standard autoregressive model definition:

\begin{equation}\label{eq:autoreg}
	y_{t+p} = \alpha_0 + \sum_{j=1}^M \alpha_j y_{t-j}
\end{equation}

\noindent where $M$ denotes the number of autoregressive terms, and the $\alpha$s are the model coefficients. This is a similar model to the one used in \cite{lazer2014parable}.

To make predictions using both web search data and historic antimicrobial drug consumption data, we combine the two above models into a single model closely resembling the approach described in \cite{yang2015accurate}. The model is as follows:

\begin{equation}\label{eq:webautoreg}
	 y_{t+p} = \theta_0 + \sum_{j=1}^M \theta_j y_{t-j} + \sum_{i=1}^N \theta_{i+M} Q_{t,i},
\end{equation}

\noindent where the $\theta$s are the model coefficients, and the remaining notation is as defined above.

For all the linear models we use Elastic Net regularization for estimating the model coefficients, as described in Equation \ref{eq:elasticnet}. The two hyper-parameters $\lambda_1$ and $\lambda_2$ are found using three fold cross-validation on the training data.

\subsubsection{Gaussian Processes for antimicrobial drug consumption prediction}
\label{sec:gp}

Gaussian Processes (GP) are probability distributions over functions, where any finite set of function values have a joint Gaussian distribution \cite{Rasmussen2004}. We focus on GP that learn functions that map from our input space of size $m$ to a single valued output, i.e. $f:R^m \rightarrow R$. The size of the input space is defined by either the number of queries, autoregressive terms, or a combination of the two.

The functions drawn from a GP can be described by two functions: A mean function and a covariance function. The mean function is defined as:

\begin{equation}
	E[f(x)] = \mu(x)
\end{equation}

\noindent where $x$ is our input data, and $\mu$ denotes the mean of the function distribution at point $x$. The covariance function is defined as:

\begin{equation}
	Cov[f(x), f(x')] = k(x, x')
\end{equation}

\noindent where $x$ and $x'$ are two input vectors, and $k$ is a kernel function \cite{Rasmussen2004}.  When working with GP it is customary to assume that the mean value is zero, and focus only on the covariance/kernel. The covariance function defines prior covariance between two input values and is typically controlled using two parameters: length scale and variance. For the Squared Exponential covariance function that we use, the variance defines the average distance from the mean, and the length scale defines how quickly the underlying signal changes, i.e. the antimicrobial drug consumption. Given the input data and the covariance function, it is possible to automatically infer the optimal parameters of the model given the data. A feature of GP is that covariance functions can be combined. This property can be used to create new covariance functions that can capture several aspects of the data. For example, combining covariance functions with different length scales could be used to model slow changes and quick changes. We use this property below.

We use two different setups, one for experiments involving web search data, and one when only historical antimicrobials data is used. The covariance function we use for web data is the Matern covariance function which allows for adapting to non-smooth changes by varying the parameter $\nu$, and is defined as:

\begin{equation}
	k_m^\nu (x, x') = \sigma^2 \frac{2^{1-\nu}}{\Gamma(\nu)} \left( \frac{\sqrt{2\nu} r}{l} \right)^\nu K_\nu \left( \frac{\sqrt{2\nu} r}{l} \right),
\end{equation}

\noindent where $\nu$ is a parameter that in our case is set to $3/2$ (a common choice), $r$ is $|x-x'|$, $l$ is the length scale, $\sigma^2$ is the variance, and $K_\nu$ is a modified Bessel function \cite{Rasmussen2004}.  To model many different types of behaviour we use the additive properties of the covariance function and generate a new covariance function, $k_{web}$, for the web search data consisting of 10 Matern functions:

\begin{equation}
	k_{web} (x, x') = \sum_{i=1}^{10} k_m^{\nu=3/2} (x, x'; \sigma_i, l_i)+N(\sigma_{11}),
\end{equation}

\noindent where $N(\sigma_{11})$ is Gaussian distributed noise.

When only working with historical antimicrobial drug consumption data, we use two other covariance functions: the linear and the squared exponential (SE). The linear covariance function can capture upwards or downwards trends in the data, and the SE function can capture short-term temporal variations in the data. The SE covariance function is defined as:

\begin{equation}
	k_{SE} (x, x') = \sigma^2 exp\left( \frac{-(x-x')^2}{2l^2} \right),
\end{equation}

\noindent where $l$ is the length scale, and $\sigma^2$ the variance. The linear covariance function is defined as:

\begin{equation}
	k_{lin}(x, x') = \sigma^2 x^Tx',
\end{equation}

\noindent with $\sigma^2$ as variance.  Combing the two covariance functions we get a new
covariance function that we use for the antimicrobial drug consumption data. We denote the covariance function $k_{antimicrobial}$ and define it as follows:

\begin{equation}
	k_{antimicrobial}(x, x') = k_{SE} (x, x'; \sigma_1, l_1) + k_{lin}(x, x'; \sigma_2)+N(\sigma_3)
\end{equation}

Parameters for all models are found using gradient descent. Different covariance functions have been tested, and we report the results of the setup with the lowest error.

\section{Experimental evaluation}

\subsection{Experimental setup}

To simulate a real world prediction situation we test the models in a leave-one-out fashion, where we re-train the prediction model after each time step such that all available data is used. This is a common setup in health event prediction \cite{yang2015accurate, www17, lampos2015advances}.

The number of autoregressive antimicrobial drug consumption terms varies between 4, 26 and 130 weeks, i.e. $M$ in Equation \ref{eq:autoreg} and \ref{eq:webautoreg}. For the web search data we generate queries with a maximum lag of 4, 26 and 130 weeks, with the specific lags selected as described in Section \ref{sec:feature_selection}. Four different prediction offsets are tested: 0, 4, 8, and 12 weeks. The prediction offset denotes how far into the future we are predicting. For example, an offset of 0 means that antimicrobial drug consumption in week $t$ is predicted using web data and historical antimicrobial drug consumption data from weeks prior to $t$. For an offset of 4 the antimicrobial drug consumption in week $t+4$ is predicted using web data and historical antimicrobial drug consumption data from weeks prior to $t$.

The web search data covers 2 January 2011 – 23 October 2016, in total 304 weeks of data. Queries are selected using the first 104 weeks of data. Each experiment uses as a minimum 104 weeks of training data for model fitting. 
With the 12 weeks of prediction offset and up 130 weeks of autoregressive terms, we end up with an evaluation period of the 58 weeks leading up to 23 October 2016.


We evaluate predictions using the root mean squared error (RMSE) and mean absolute error (MAE). A feature of the RMSE is that large prediction errors receive a bigger penalty than small errors. This intuitively means that few large errors will result in a larger RMSE than many small errors. The MAE, on the other hand, assigns equal weight to all errors, and the final score is therefore easier to interpret. With respect to our data, a MAE of 10000 corresponds to the prediction on average being 10000 DDDs off on every weekly prediction. This corresponds to approximately 6\% of the weekly average of DDDs.

The RMSE is calculated as:

\begin{equation}
	RMSE = \sqrt{1/N \sum_{t=1}^N(y_t - \hat{y}_t)^2}
\end{equation}

\noindent where $y_t$ is the true value at time $t$, $\hat{y}_t$ is the predicted value at time $t$, and $N$ the number of predictions. The MAE is calculated as:

\begin{equation}
	MAE = 1/N \sum_{t=1}^N |y_t - \hat{y}_t|.
\end{equation}


\subsection{Experimental results}



\begin{table}
	\centering
	\scalebox{0.9}{
	\begin{tabular}{p{1.1cm}lrr}
		\toprule
		Offset    & Data Source        &      GP & Elastic Net \\ \midrule
		\multirow{3}{*}{0 weeks}  & Web only           & 11011.0 &     11096.9 \\
		                    & Antimicrobial only & 11446.0 &      9980.9 \\
		                    & Web \& antimicrobial           & 10644.3 &      \textbf{9970.8}\vspace{0.1cm} \\
		\multirow{3}{*}{4 weeks}  & Web only           & 11270.7 &     11398.1 \\
		                    & Antimicrobial only & 11498.2 &      9990.3 \\
		                    & Web \& antimicrobial           & 10576.2 &      \textbf{9989.9}\vspace{0.1cm} \\
		\multirow{3}{*}{8 weeks}  & Web only           & 11026.1 &     11142.5 \\
		                    & Antimicrobial only & 11401.6 &     10301.0 \\
		                    & Web \& antimicrobial           & 10564.4 &      \textbf{9781.0}\vspace{0.1cm} \\
		\multirow{3}{*}{12 weeks} & Web only           & 10977.8 &     11189.1 \\
		                    & Antimicrobial only & 11231.0 &     10424.0 \\
		                    & Web \& antimicrobial           & 10249.1 &      \textbf{9644.3} \\ \bottomrule
	\end{tabular}}
	\caption{RMSE for best performing prediction (among all query sets) with the non-linear (GP) and linear (Elastic Net) prediction model, using web, antimicrobial purchase data, and their combination. Lowest error per offset is in bold.}
	\label{tabel:rmse}
\end{table}

\begin{table}
	\centering
	\scalebox{0.9}{
	\begin{tabular}{p{1.1cm}lrr}
		\toprule
		Offset      & Data Source        &     GP & Elastic Net \\ \midrule
		\multirow{3}{*}{0 weeks}  & Web only           & 8463.4 &      8507.7 \\
		                    & Antimicrobial only & 8571.0 &      7294.5 \\
		                    & Web \& antimicrobial            & 8105.3 &      \textbf{7282.7}\vspace{0.1cm} \\
		\multirow{3}{*}{4 weeks}  & Web only           & 8938.2 &      9040.8 \\
		                    & Antimicrobial only & 9265.8 &      \textbf{7989.9} \\
		                    & Web \& antimicrobial            & 8440.0 &      8073.2\vspace{0.1cm} \\
		\multirow{3}{*}{8 weeks}  & Web only           & 8596.4 &      8579.8 \\
		                    & Antimicrobial only & 9053.2 &      8200.7 \\
		                    & Web \& antimicrobial            & 8311.2 &      \textbf{7849.9}\vspace{0.1cm} \\
		\multirow{3}{*}{12 weeks} & Web only           & 8258.8 &      8463.5 \\
		                    & Antimicrobial only & 8997.7 &      8257.6 \\
		                    & Web \& antimicrobial            & 8056.7 &      \textbf{7750.0} \\ \bottomrule
	\end{tabular}
}
	\caption{MAE for best performing prediction (among all query sets) with the non-linear (GP) and linear (Elastic Net) prediction model, using web, antimicrobial purchase data, and their combination. Lowest error per offset is in bold.}
	\label{tabel:mae}
\end{table}

\begin{table}
	\centering
	\scalebox{0.9}{
	\begin{tabular}{p{1.1cm}lRR}
		\toprule
		Offset                    & Data Source          &    GP &                  Elastic Net \\ \midrule
		\multirow{3}{*}{0 weeks}  & Web only             & 5.4\% &                        5.4\% \\
		                          & Antimicrobial only   & 5.4\% &               \textbf{4.6}\% \\
		                          & Web \& antimicrobial & 5.1\% & \textbf{4.6}\%\vspace{0.1cm} \\
		\multirow{3}{*}{4 weeks}  & Web only             & 5.7\% &                        5.7\% \\
		                          & Antimicrobial only   & 5.9\% &               \textbf{5.1}\% \\
		                          & Web \& antimicrobial & 5.4\% & \textbf{5.1}\%\vspace{0.1cm} \\
		\multirow{3}{*}{8 weeks}  & Web only             & 5.5\% &                        5.5\% \\
		                          & Antimicrobial only   & 5.8\% &                        5.2\% \\
		                          & Web \& antimicrobial & 5.3\% & \textbf{5.0}\%\vspace{0.1cm} \\
		\multirow{3}{*}{12 weeks} & Web only             & 5.2\% &                        5.4\% \\
		                          & Antimicrobial only   & 5.7\% &                        5.2\% \\
		                          & Web \& antimicrobial & 5.1\% &               \textbf{4.9}\% \\ \bottomrule
	\end{tabular}}
	\caption{MAE as a percentage of the average weekly antimicrobial usage for the 58 week evaluation period.}
	\label{tabel:mae_percentage}
\end{table}

\begin{figure}
	\centering
	\includegraphics[width=\linewidth]{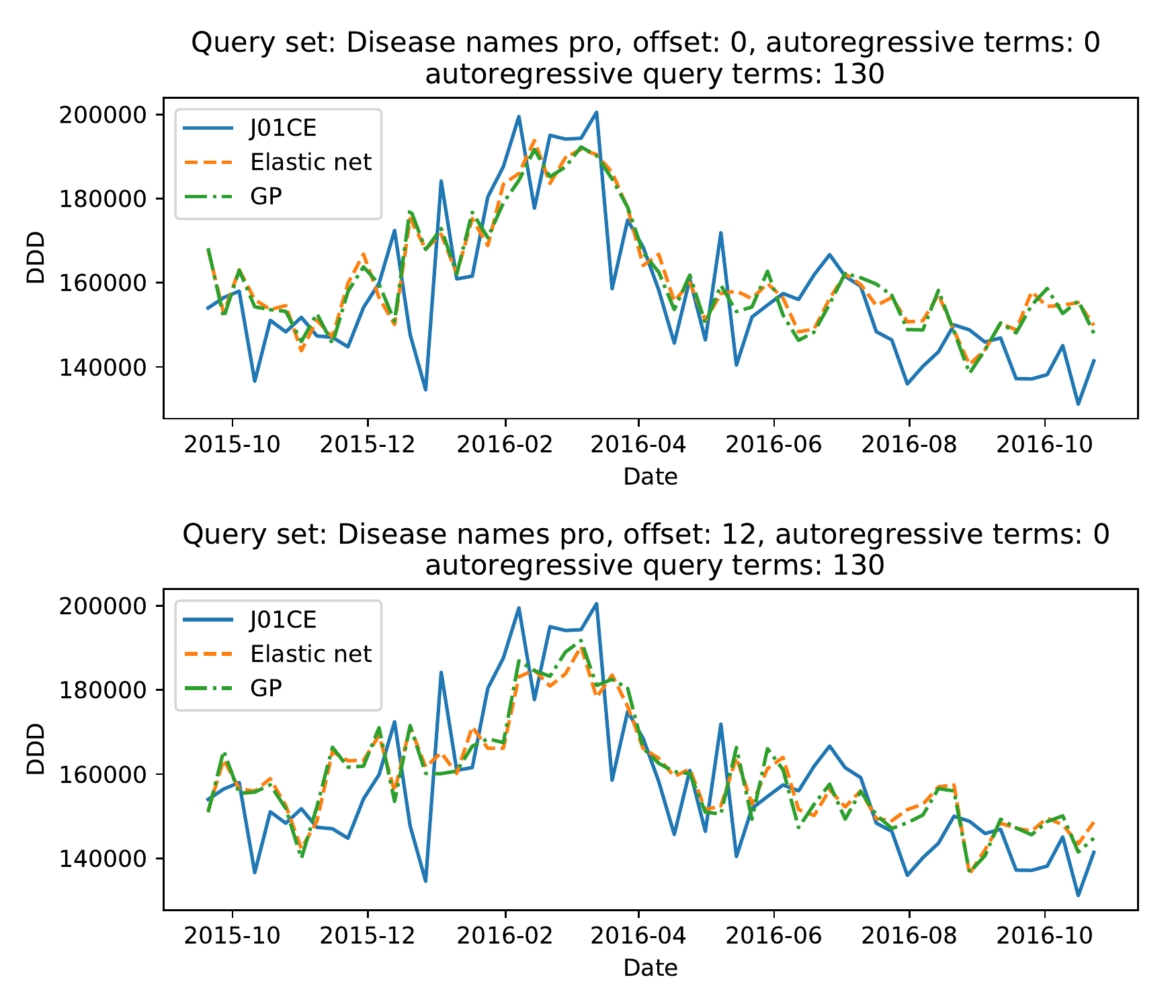}
	\caption{Prediction using only web search data.}
	\label{fig:prediction_web}
\end{figure}

\begin{figure}
	\centering
	\includegraphics[width=\linewidth]{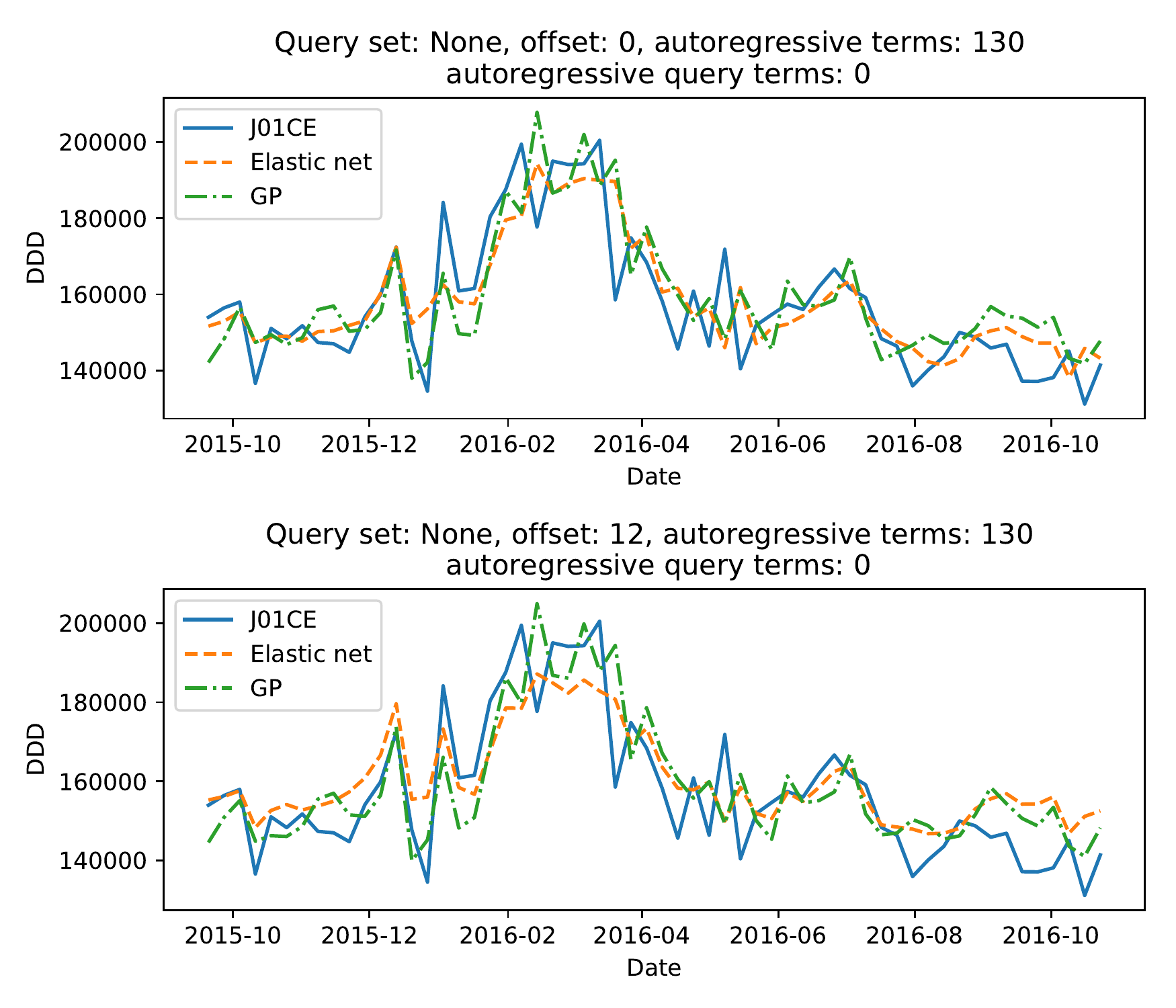}
	\caption{Prediction using only historic antimicrobial purchase data.}
	\label{fig:prediction_antimicrobial}
\end{figure}

\begin{figure}
	\centering
	\includegraphics[width=\linewidth]{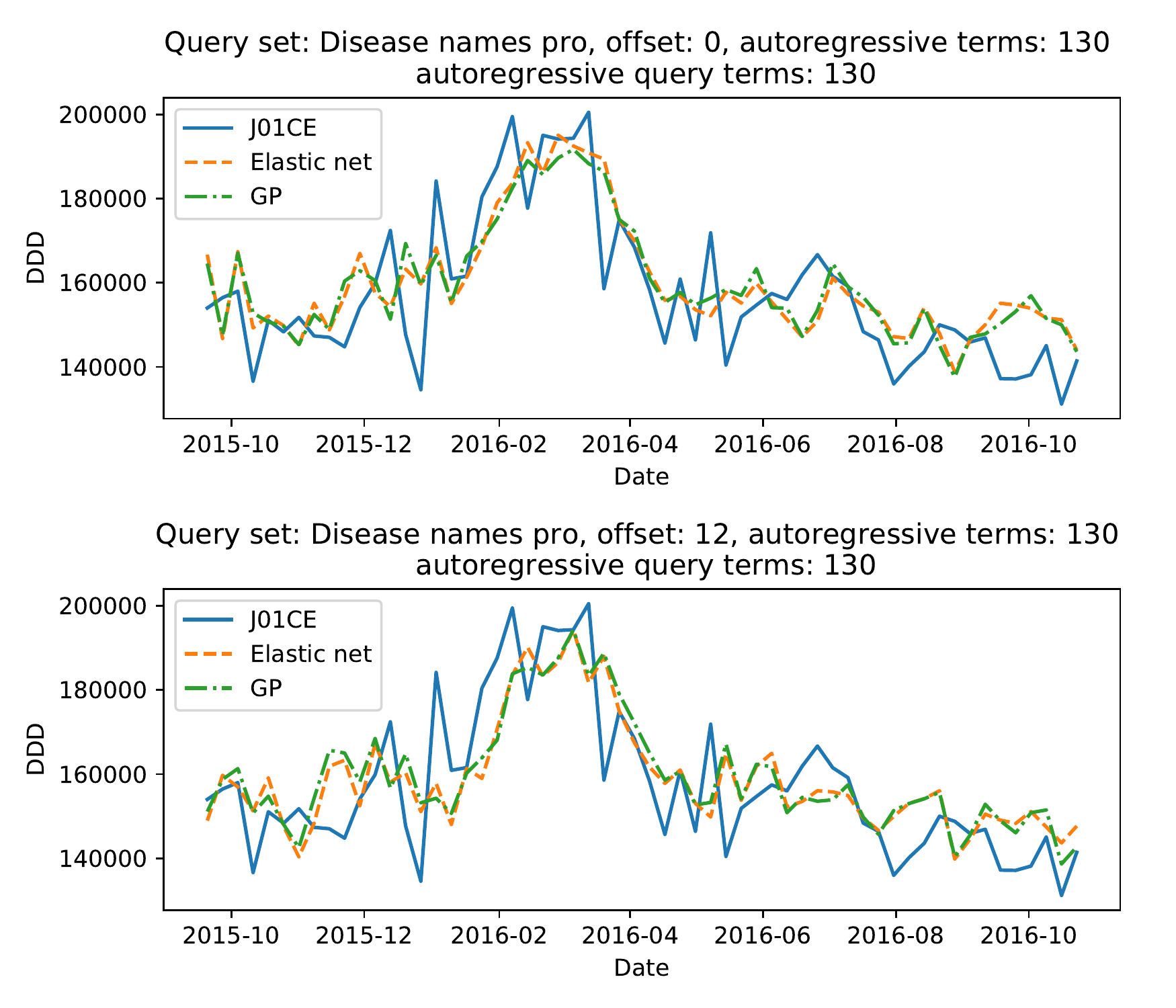}
	\caption{Prediction using both web and antimicrobial purchase data.}
	\label{fig:prediction_combined}
\end{figure}

We start by presenting the results with respect to model selection and prediction offset. Subsequently in Section \ref{sec:query_set_performance} we discuss the influence of query set and time lag selection.

Tables \ref{tabel:rmse} \& \ref{tabel:mae} show the RMSE and MAE when predicting antimicrobial drug consumption using (i) only web search data, (ii) only historical antimicrobial purchase data, and (iii) both web search and antimicrobial purchase data. Only the best performance (lowest error) is reported per data source, prediction model, and offset.

We see that predictions based on a combination of web and antimicrobial purchase data give almost always the lowest error. With Gaussian Processes as the prediction model, predictions based only on web data outperform those based on antimicrobial purchase data; with Elastic Net, the situation is reversed: predictions based on antimicrobial purchase data outperform those based on web data.
Comparing the two prediction models (Gaussian Processes and Elastic Net), we only observe minor differences, generally favouring the linear models. These results fit well with the general prevalence of linear models for prediction using web search data, though it is curious why Gaussian Processes in other work outperform Elastic Net on similar tasks \cite{www17_lampos, lampos2015advances}. One explanation could be that the feature selection described in Section \ref{sec:feature_selection} uses Elastic Net, which means that we are selecting features that have a linear relationship with the target variable, i.e. antimicrobial drug consumption. In other words, we have a priori selected features that are well suited for the Elastic Net model.

Overall, the fluctuations in error (both RMSE and MAE) across different prediction models, data sources, and offsets are generally small. This is also illustrated in Table \ref{tabel:mae_percentage}, which shows the MAE scores of Table \ref{tabel:mae} as the \% of average weekly consumption in the evaluation period.  We see that our predictions are, in the best case, off by 4.6\% of the average weekly consumption, and in the worst case by 5.9\%.

We also observe in Tables  \ref{tabel:rmse} -- \ref{tabel:mae_percentage} that error remains generally stable independent of offset. Looking back at Figure \ref{fig:j0ce_usage} we observe two things which might explain this: (i) The antimicrobial  drug consumption is strongly seasonal, therefore models that capture the latent seasonality will perform well even with a large offset. (ii) As will be described next, many of the queries with highest model coefficients have approximately 1 year lag. Combined with the fact that the consumption patterns in the last three years of the time series are very similar, we should expect to be able to predict the antimicrobial drug consumption relatively accurately one year into the future.


In Figures \ref{fig:prediction_web} -- \ref{fig:prediction_antimicrobial} we further plot the predictions by the two models (GP and Elastic Net) using only web or only antimicrobial purchase data against actual antimicrobial purchase data (J01CE). The precise settings of these four runs are stated in the figure titles. 
We see that, when using web data only, seasonal variations are generally captured; however, a drop in antimicrobial drug consumption in January 2016 is not captured. Visually, the difference between a 0 week offset and a 12 week offset is hard to spot. When using only antimicrobial purchase data, on the other hand, the GP model captures the drop in consumption in January 2016, both with a 0 week offset and a 12 week offset.  Again differences between 0 week offset and 12 week offset are negligible. Finally, Figure \ref{fig:prediction_combined} shows the combination of the two data sources. Here neither model captures the drop in January 2016. Differences between 0 week offset and 12 week offset are, as before, minor.

Overall, we find that the use of web data only gives predictions that are slightly more erroneous, but generally not that far off, from those made when using only historical antimicrobial purchase data. For both types of data we find that long term variations are consistently captured, while precise short term predictions, e.g. drop in consumption in January 2016 Figure \ref{fig:prediction_antimicrobial}, are better captured using historic antimicrobial data. As a tool for maintaining political focus and analyzing general usage patterns, short term precision is likely of less importance. The fact that the difference between the data sources is so small is valuable for countries lacking timely access to centralised antimicrobial purchase data, because it means that we can approximate predictions that are roughly less than 1\% point erroneous compared to those using antimicrobial purchase data (for the same offset -- cf. Table \ref{tabel:mae_percentage}).  This performance appears generally stable across different prediction offsets and linear (Elastic Net) vs non-linear (GP) prediction models.

Next we analyse the impact of web search query selection to prediction performance.

\subsubsection{Web search query analysis}
\label{sec:query_set_performance}

\begin{figure*}[!h]
	\begin{subfigure}[b]{0.32\textwidth}
		\includegraphics[width=\linewidth]{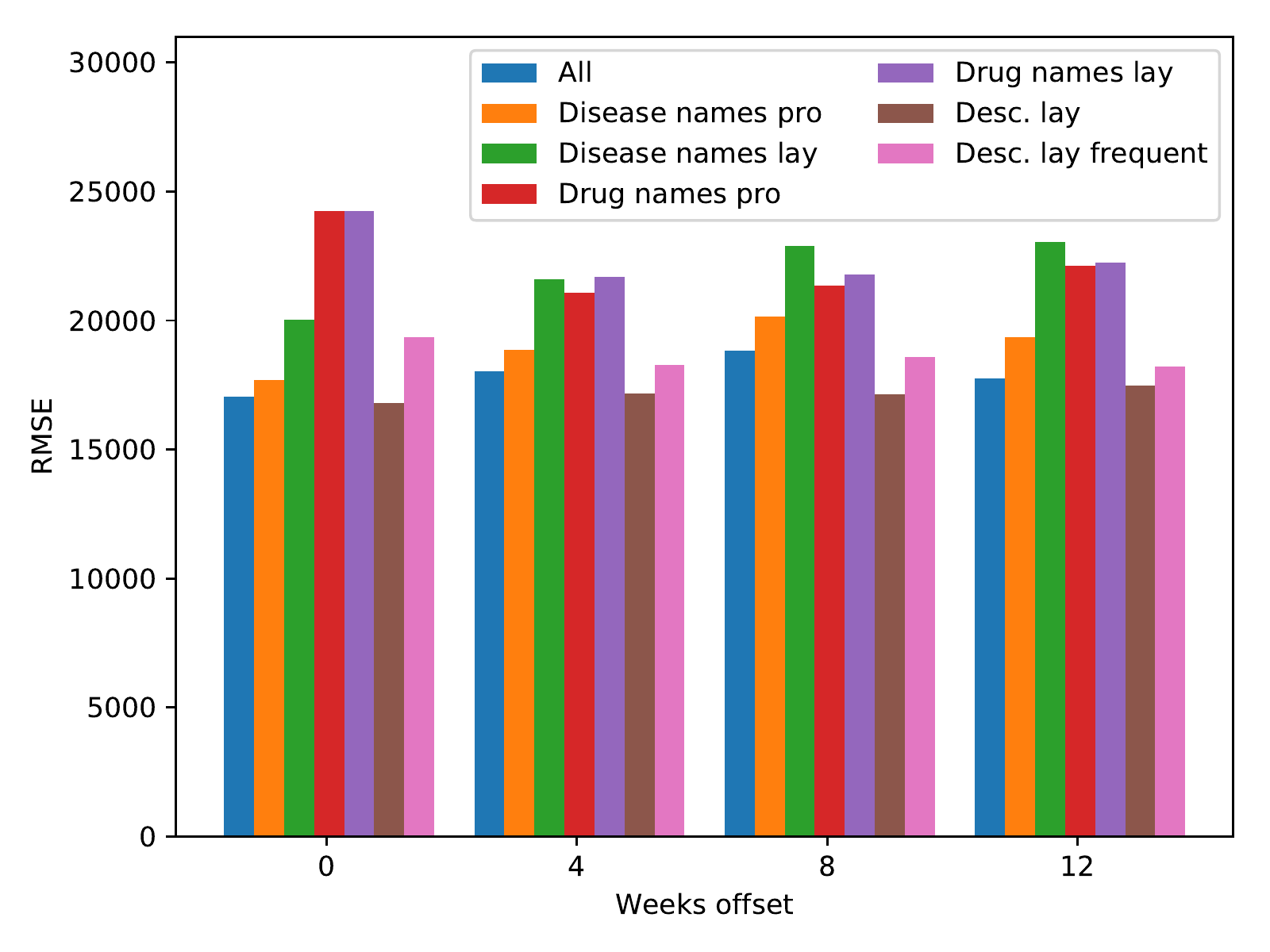}
		\caption{Up to 4 weeks lag}
	\end{subfigure}
	\begin{subfigure}[b]{0.32\textwidth}
		\includegraphics[width=\linewidth]{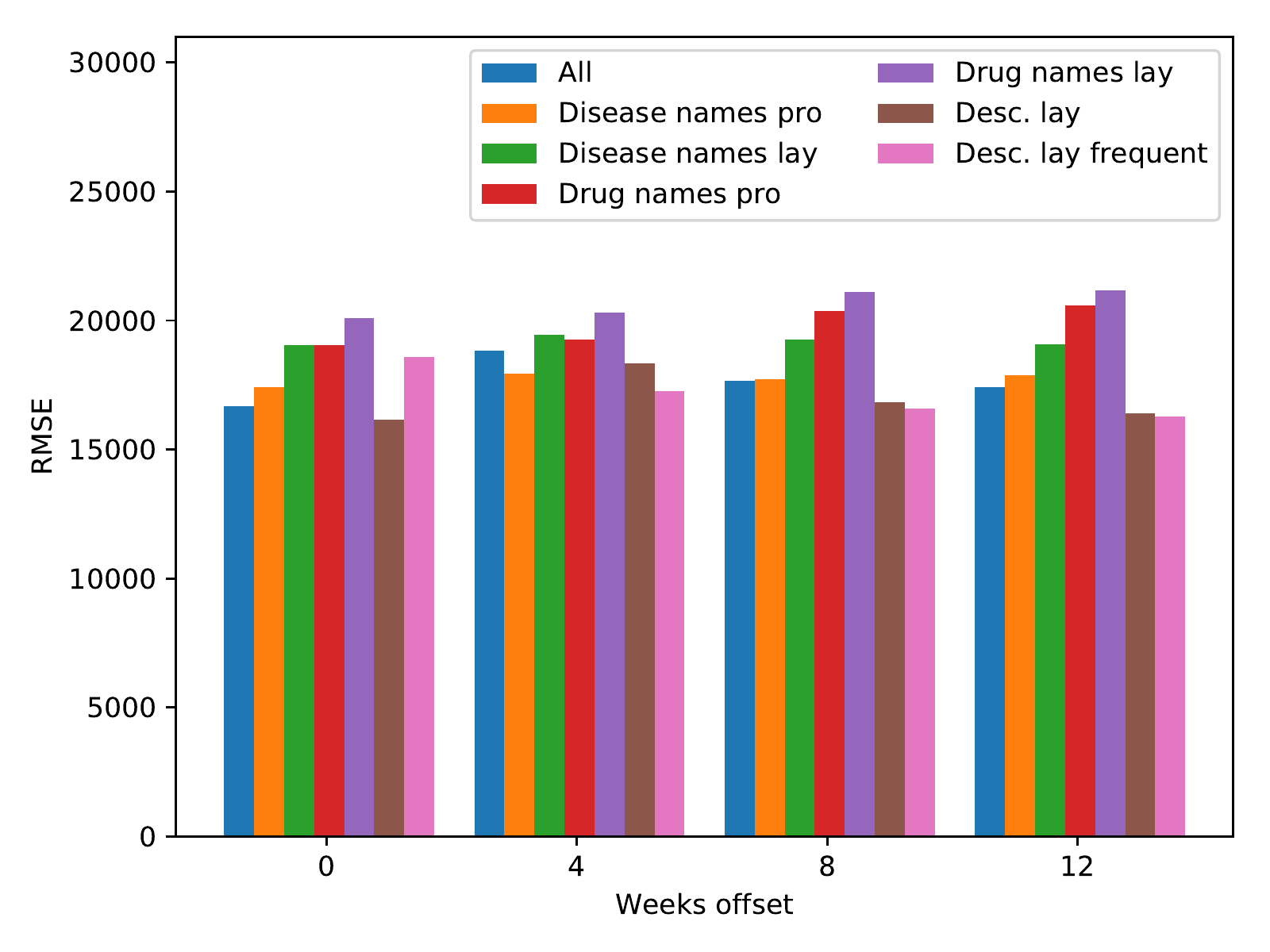}
		\caption{Up to 26 weeks lag}
	\end{subfigure}
	\begin{subfigure}[b]{0.32\textwidth}
		\includegraphics[width=\linewidth]{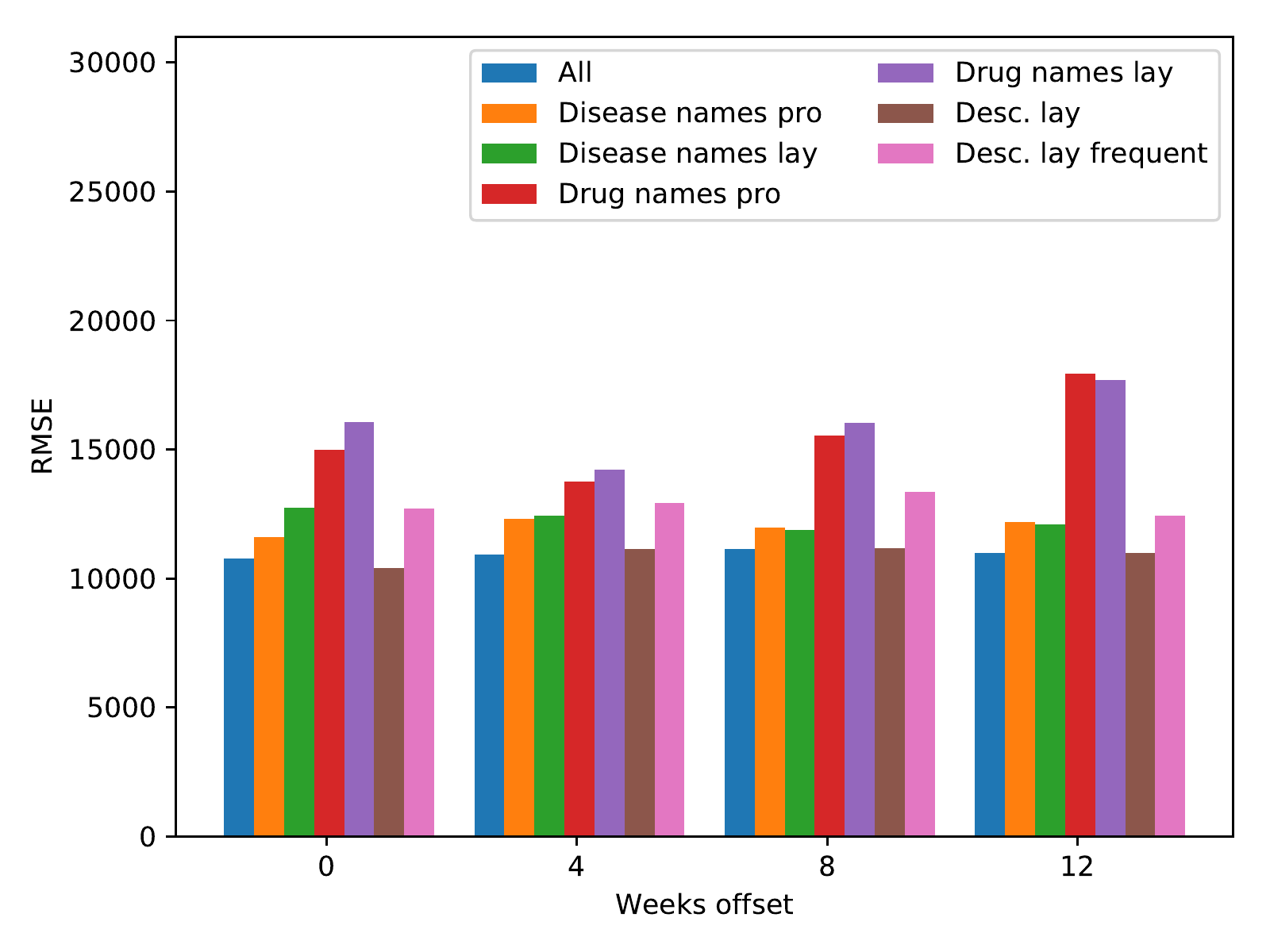}
		\caption{Up to 130 weeks lag}
	\end{subfigure}
	\caption{Prediction based on web search data with different query sets and lags, using Gaussian Processes.}
	\label{fig:query_set_performance_web-gp}

\end{figure*}
\begin{figure*}[!h]

	\begin{subfigure}[b]{0.32\textwidth}
		\includegraphics[width=\linewidth]{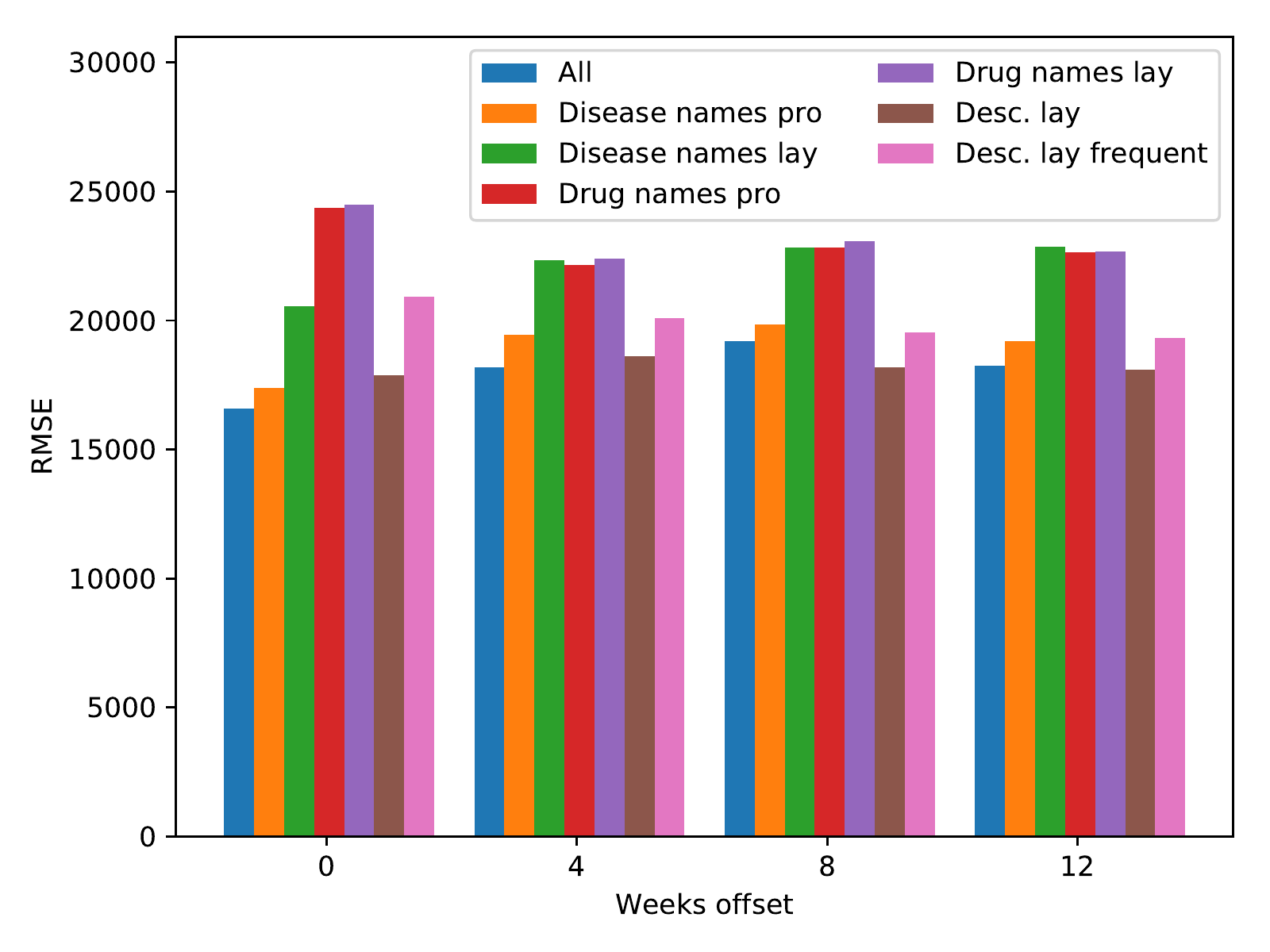}
		\caption{Up to 4 weeks lag}
	\end{subfigure}
	\begin{subfigure}[b]{0.32\textwidth}
		\includegraphics[width=\linewidth]{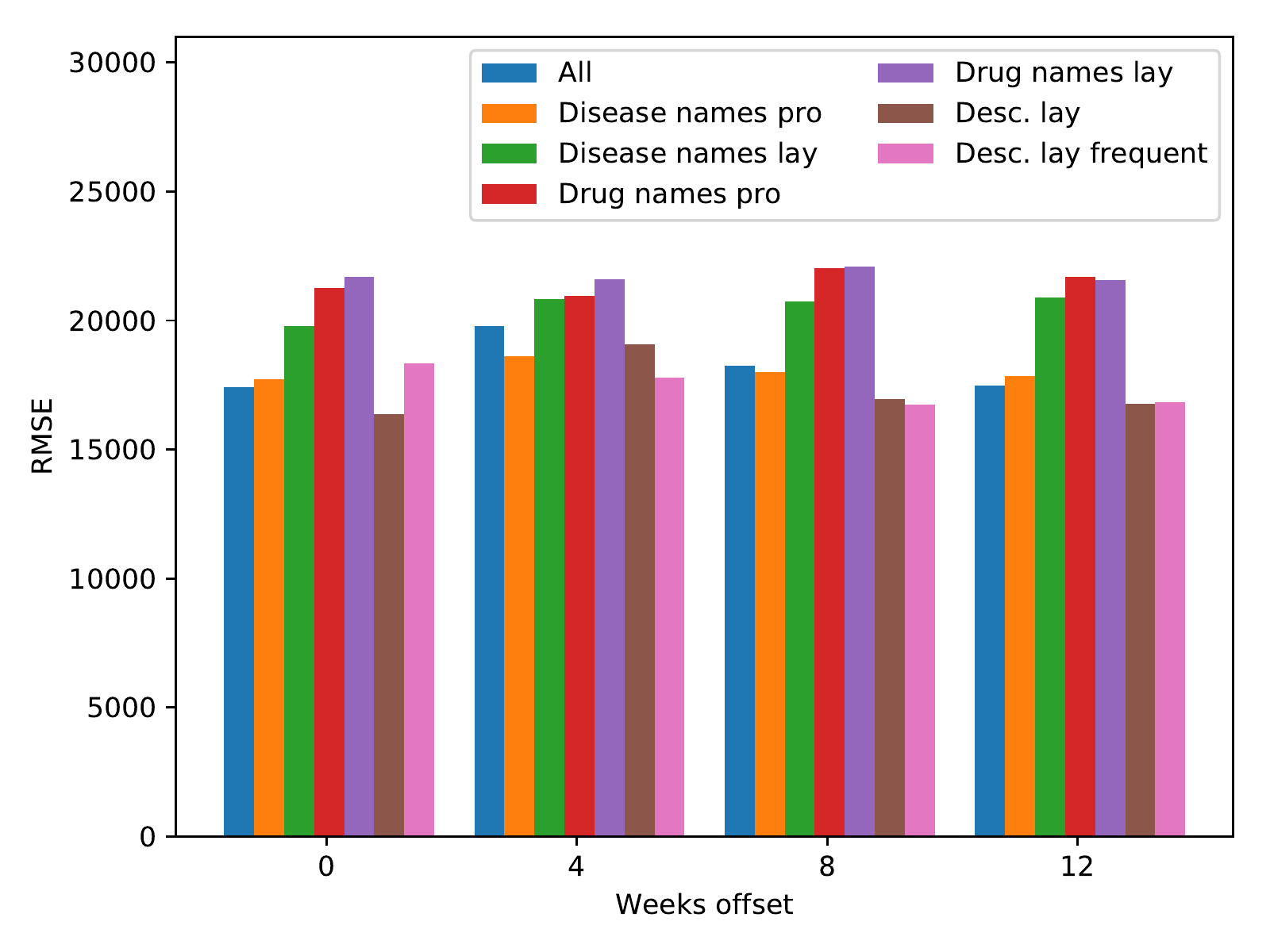}
		\caption{Up to 26 weeks lag}
	\end{subfigure}
	\begin{subfigure}[b]{0.32\textwidth}
		\includegraphics[width=\linewidth]{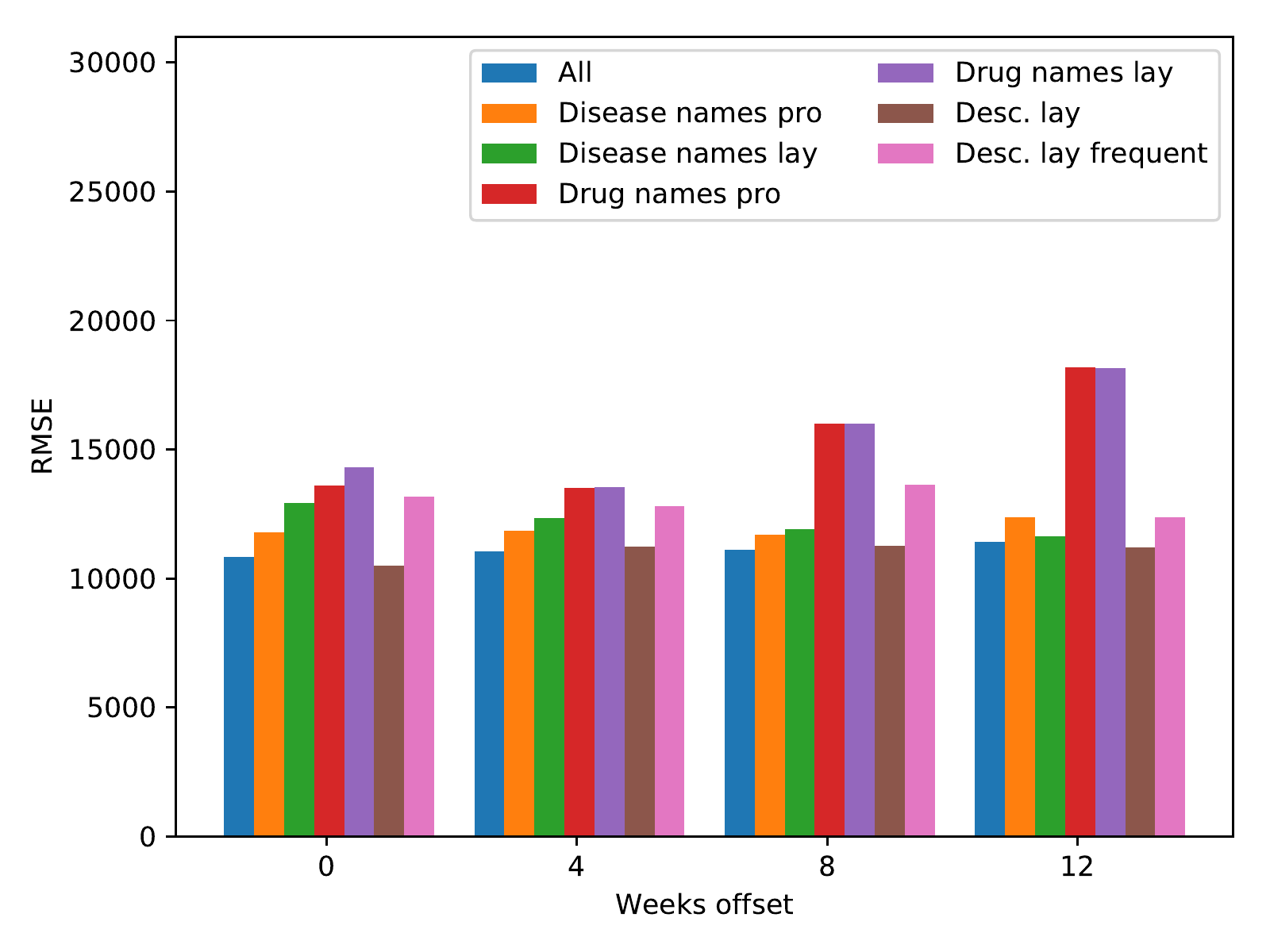}
		\caption{Up to 130 weeks lag}
	\end{subfigure}
	\caption{Prediction based on web search data with different query sets and lags, using Elastic Net.}
	\label{fig:query_set_performance_web-en}

\end{figure*}
\begin{figure*}[!h]

	\begin{subfigure}[b]{0.32\textwidth}
		\includegraphics[width=\linewidth]{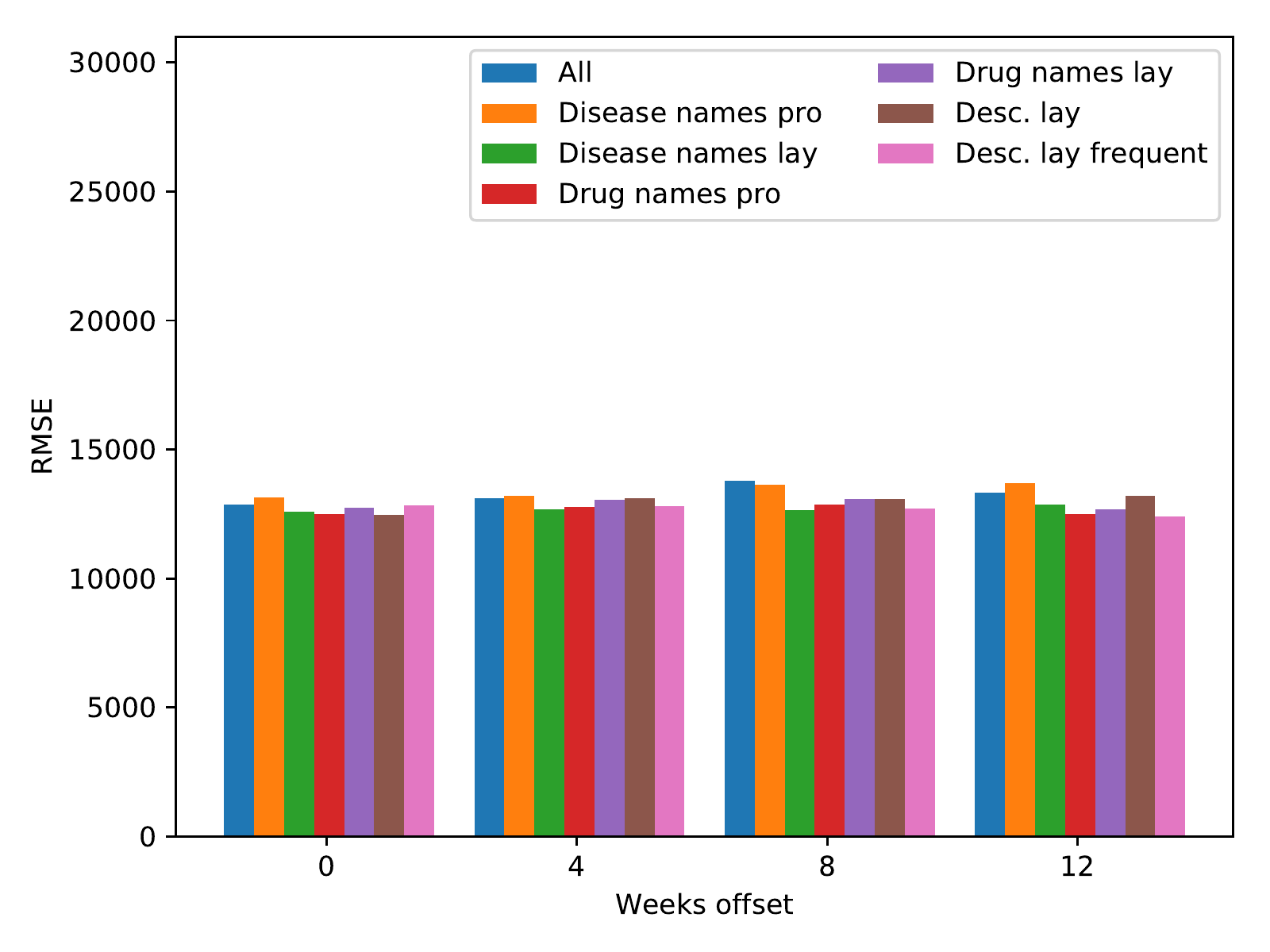}
		\caption{Up to 4 weeks lag}
	\end{subfigure}
	\begin{subfigure}[b]{0.32\textwidth}
		\includegraphics[width=\linewidth]{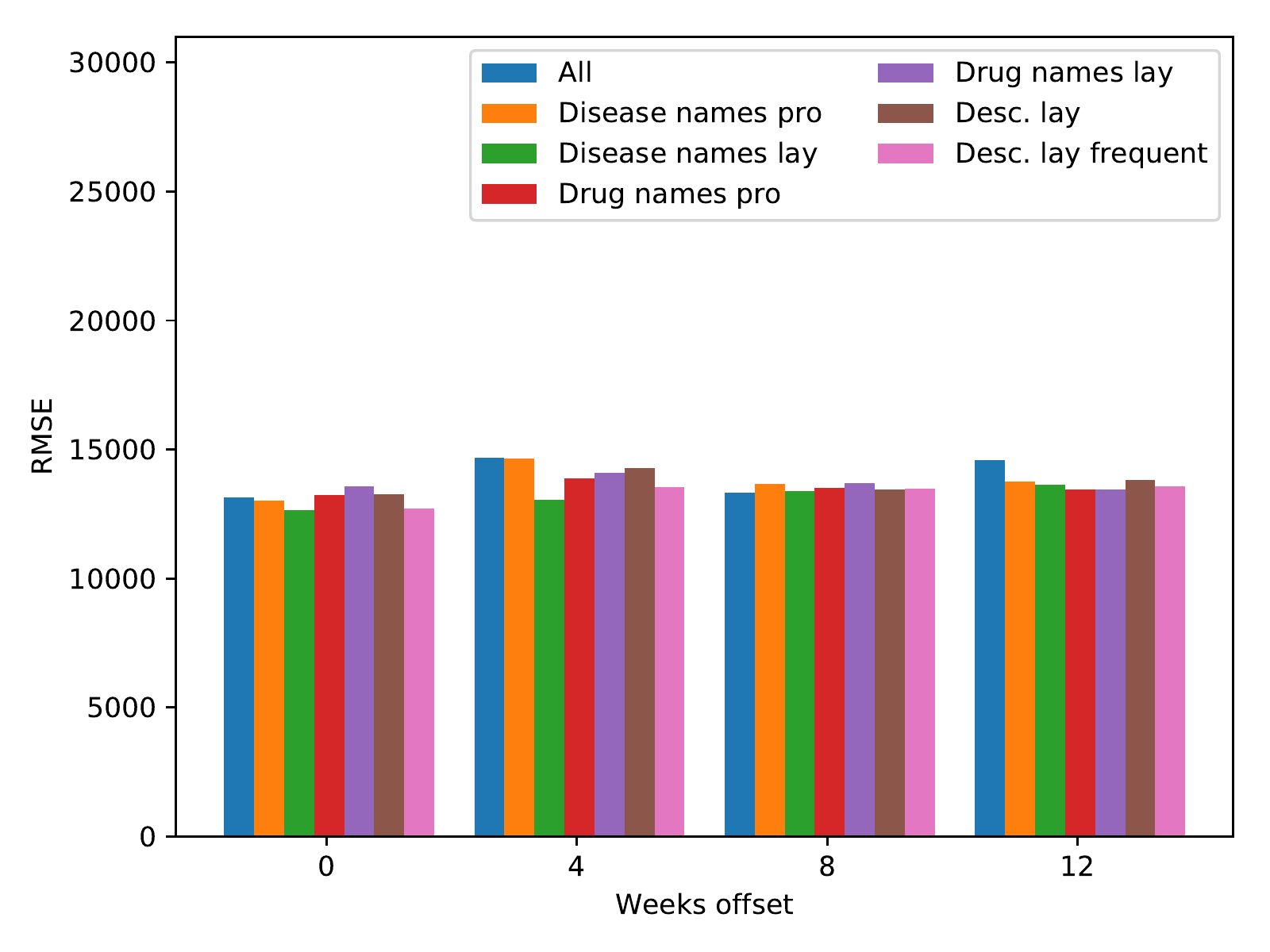}
		\caption{Up to 26 weeks lag}
	\end{subfigure}
	\begin{subfigure}[b]{0.32\textwidth}
		\includegraphics[width=\linewidth]{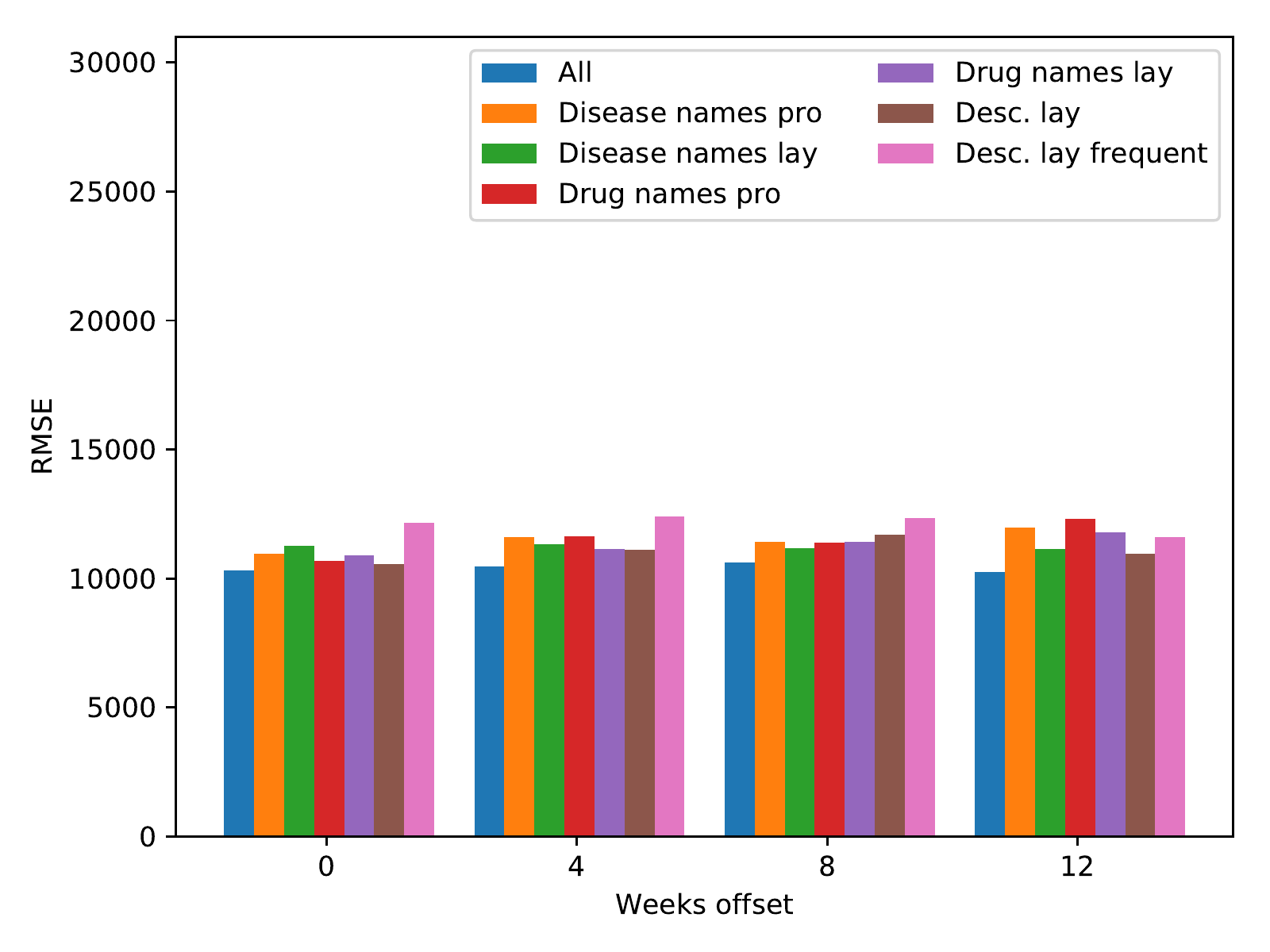}
		\caption{Up to 130 weeks lag}
	\end{subfigure}
	\caption{Prediction based on both web data and antimicrobial purchase data (130 autoregressive terms), with different query sets and lags, using Gaussian Processes.}
	\label{fig:query_set_performance_antibiotis-gp}

\end{figure*}
\begin{figure*}[!h]

	\begin{subfigure}[b]{0.32\textwidth}
		\includegraphics[width=\linewidth]{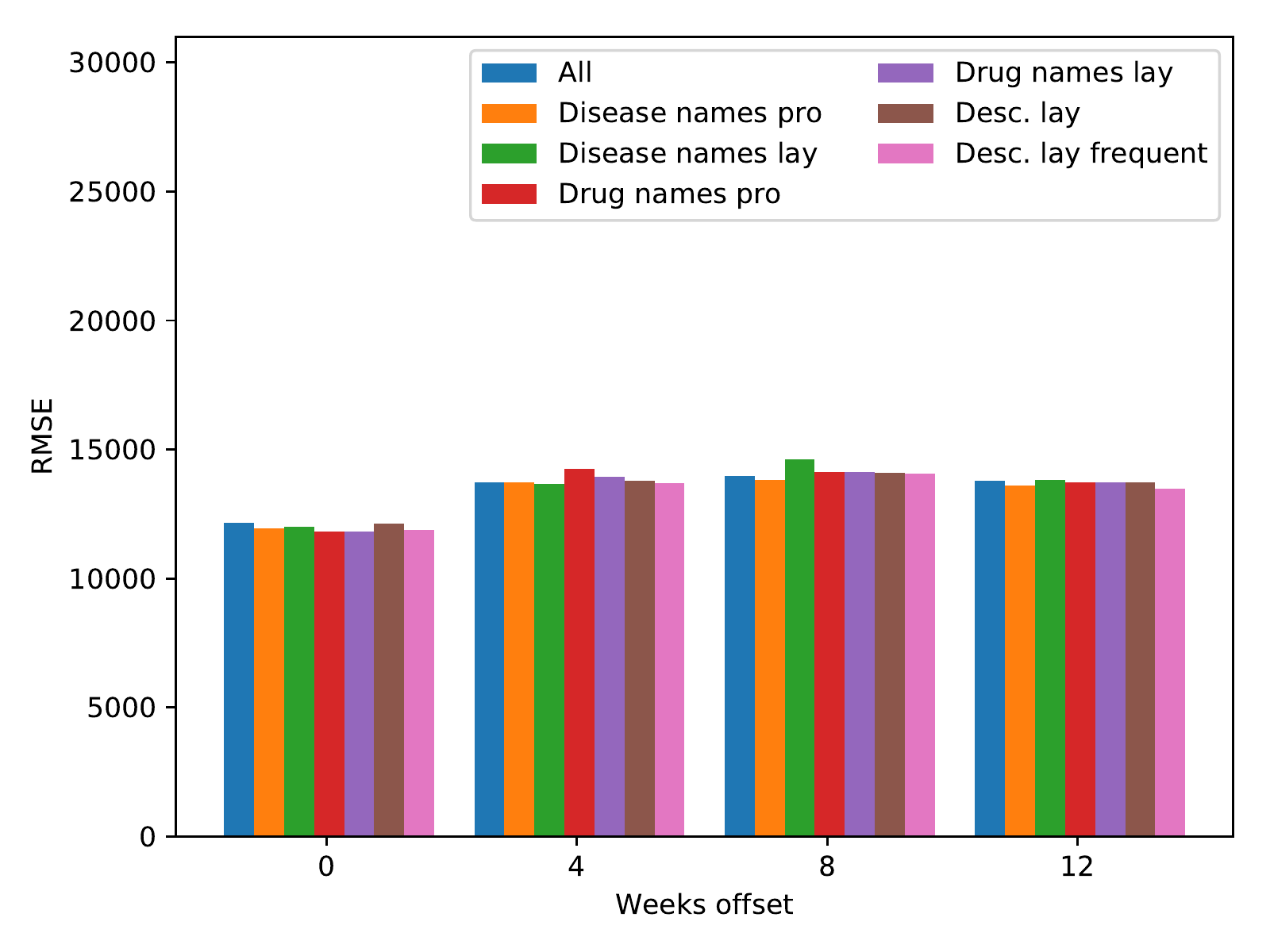}
		\caption{Up to 4 weeks lag}
	\end{subfigure}
	\begin{subfigure}[b]{0.32\textwidth}
		\includegraphics[width=\linewidth]{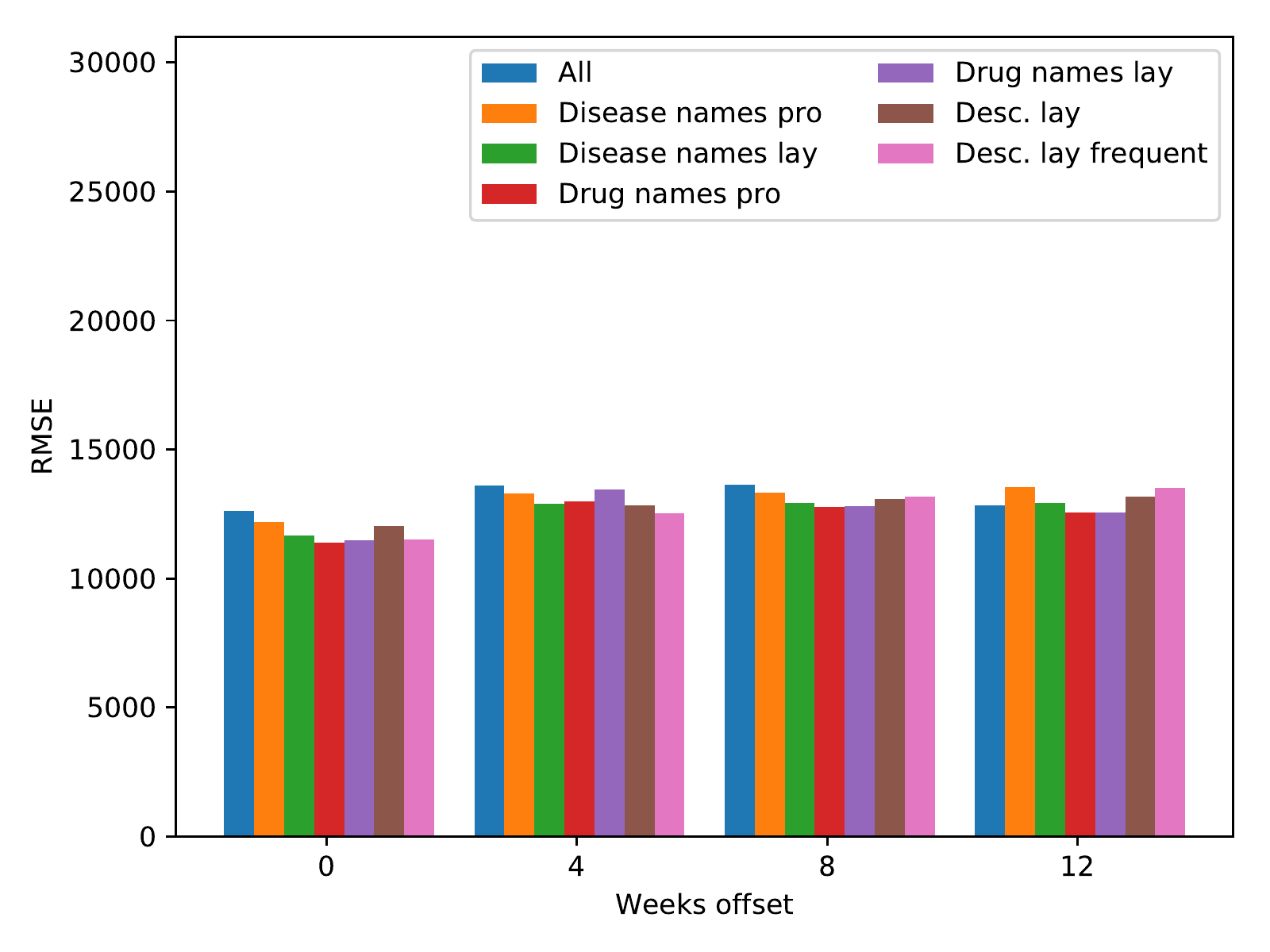}
		\caption{Up to 26 weeks lag}
	\end{subfigure}
	\begin{subfigure}[b]{0.32\textwidth}
		\includegraphics[width=\linewidth]{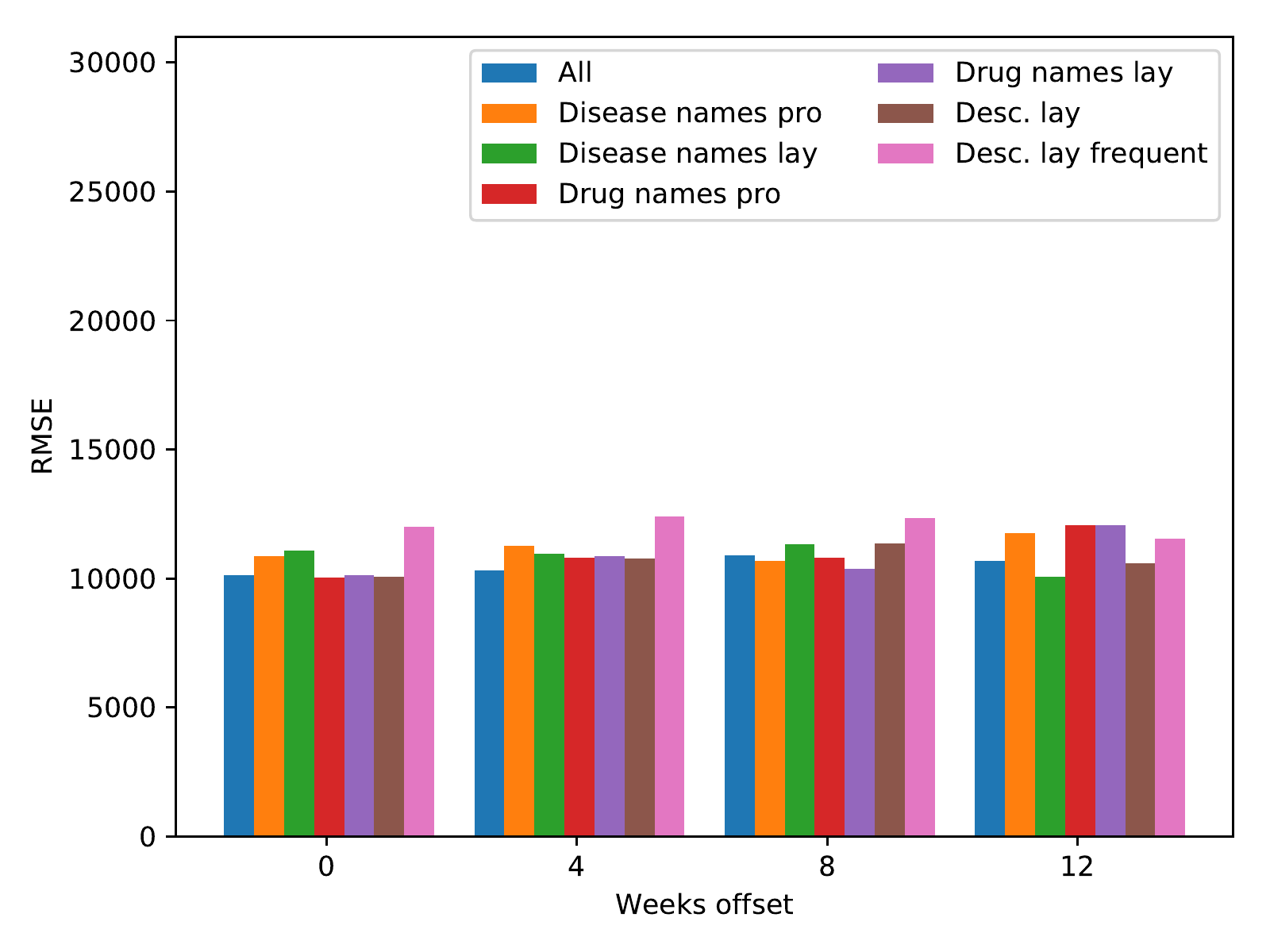}
		\caption{Up to 130 weeks lag}
	\end{subfigure}
	\caption{Prediction based on both web data and antimicrobial purchase data (130 autoregressive terms), with different query sets and lags, using Elastic Net.}
	\label{fig:query_set_performance_antibiotis-en}

\end{figure*}

Depending on the amount of maximum number of lags used, different queries are selected from the six query sets displayed in Table \ref{table:query_count}. We vary the maximum number of lags between 4, 26 and 130 weeks. Table \ref{tabel:top10_4weeks} shows the top 10 queries from each query set when using 4 weeks of historical antimicrobial data. The queries are selected using the method described in Section \ref{sec:feature_selection}. We see that most of the queries listed in Table \ref{tabel:top10_4weeks} are diseases curable with antimicrobials, such as Scarlet Fever and pneumonia. We also see diseases such as psoriasis, which itself is not treatable with antimicrobials, but which increases the risk of skin infections. It is interesting to note that even rare diseases, such as anthrax (typically a non-lethal skin infection) and syphilis, are in the top 10; this occurs because both of these diseases have antimicrobials as primary treatment. For the queries derived from the laymen descriptions of antimicrobials, there is a number of spurious correlations, e.g. words such as ``one'', ``effect'', etc. While these queries are apparently well correlated to antimicrobial drug consumption, they are semantically unrelated to antimicrobial usage, meaning that their generalisable discriminative strength is limited. 

\begin{table*}
	\centering
	\scalebox{0.9}{
	\begin{tabular}{lp{13.5cm}}
		\toprule
		Query set & Top 10 queries \\
		\midrule
		Disease names pro & psoriasis$_{t-2}$, skarlagensfeber$_{t-3}$  (\textit{Scarlet Fever}), skarlagensfeber$_{t-2}$, lungebetændelse$_{t-4}$ (\textit{pneumonia}), prostatitis$_{t-4}$, diabetes - type 2$_{t-3}$, psoriasis$_{t-3}$, blindtarmsbetændelse$_{t-3}$ (\textit{appendicitis}), lungebetændelse$_{t-3}$, blyforgiftning$_{t-4}$ (\textit{lead poisoning})\vspace{0.1cm} \\
		Disease names lay &skarlagensfeber$_{t-2}$, skarlagensfeber$_{t-3}$, skarlagensfeber$_{t-1}$, endokardit$_{t-2}$ (\textit{endocarditis}), endokardit$_{t-1}$, endokardit$_{t-3}$, endokardit$_{t-4}$, brystbetændelse$_{t-1}$ (\textit{mastitis}), syfilis$_{t-3}$ (\textit{syphilis}), miltbrand$_{t-2}$ (\textit{anthrax})\vspace{0.1cm}\\
		Descriptions lay & lungebetændelse$_{t-3}$, én$_{t-2}$ (\textit{one}), lungebetændelse$_{t-2}$, én$_{t-4}$, virkning$_{t-3}$ (\textit{effect}), én$_{t-1}$, lungebetændelse$_{t-4}$, halsbetændelse$_{t-1}$ (\textit{sore throat}), stoffer$_{t-1}$ (\textit{drugs}), én$_{t-3}$ \\
		\bottomrule
	\end{tabular}}
	\caption{Top 10 queries for each query set generated with a maximum lag of 4 weeks and a 0 week prediction offset. The subscript denotes the offset from the current prediction point. \textit{Drug names lay}, \textit{drug names pro} and \textit{descriptions lay frequent} have zero valued coefficients and are therefore omitted from the Table. English translations in brackets.}
	\label{tabel:top10_4weeks}

	\centering
	\scalebox{0.9}{
	\begin{tabular}{lp{13.5cm}}
		\toprule
		Query set & Top 10 queries \\
		\midrule
		Disease names pro & blindtarmsbetændelse$_{t-6}$ (\textit{appendicitis}), caries$_{t-1}$ , skarlagensfeber$_{t-52}$ (\textit{Scarlet Fever}), diabetes - type 2$_{t-121}$, kol$_{t-83}$ (\textit{COPD}), gasgangræn$_{t-38}$ (\textit{gas gangrene}), kol$_{t-3}$, caries$_{t-125}$, diabetisk neuropati$_{t-112}$, kol$_{t-82}$ \vspace{0.1cm} \\
		Disease names lay & skarlagensfeber$_{t-52}$, skarlagensfeber$_{t-53}$, skarlagensfeber$_{t-51}$, skarlagensfeber$_{t-50}$, skarlagensfeber$_{t-55}$, skarlagensfeber$_{t-54}$, gasgangræn$_{t-100}$, skarlagensfeber$_{t-3}$, gasgangræn$_{t-38}$, skarlagensfeber$_{t-103}$ \vspace{0.1cm}\\
		Drug names pro & novu$_{t-59}$, novu$_{t-111}$, novu$_{t-116}$, novu$_{t-30}$, novu$_{t-9}$, novu$_{t-62}$, novu$_{t-32}$, novu$_{t-7}$, novu$_{t-127}$, novu$_{t-82}$\vspace{0.1cm}\\
		Drug names lay & novu$_{t-59}$, novu$_{t-111}$, novu$_{t-62}$, novu$_{t-9}$, novu$_{t-30}$, novu$_{t-66}$, novu$_{t-7}$, novu$_{t-116}$, novu$_{t-33}$, novu$_{t-32}$ \vspace{0.1cm}\\
		Descriptions lay & skyldes$_{t-107}$ (\textit{due}), vækst$_{t-94}$ (\textit{growth}), udvikle$_{t-123}$ (\textit{develop}), immunforsvar$_{t-83}$ (\textit{immune system}), resistens$_{t-4}$ (\textit{resistance}), allergi$_{t-92}$ (\textit{allergy}), bivirkninger$_{t-9}$ (\textit{side-effect}), dræbe$_{t-99}$ (\textit{kill}), behandlingen$_{t-85}$ (\textit{treatment}), skyldes$_{t-104}$\vspace{0.1cm}\\
		Descriptions lay frequent & infektion$_{t-51}$ (\textit{infection}), vækst$_{t-55}$, vækst$_{t-94}$, infektion$_{t-78}$, medicin$_{t-4}$ (\textit{medicine}), virus$_{t-102}$, bakterier$_{t-83}$ (\textit{bacteria}), vækst$_{t-42}$, behandling$_{t-62}$, bakterierne$_{t-117}$ \\
		\bottomrule
	\end{tabular}}
	\caption{Top 10 queries for each query set with a maximum lag of 130 and a 0 week prediction offset. The subscript denotes the offset from the current prediction point. English translations in brackets.}
	\label{tabel:top10_130weeks}
\end{table*}

Table \ref{tabel:top10_130weeks} further shows the top 10 queries for a lag of up to 130 weeks. In this case we clearly begin to see the effect of seasonality (seen in Figure \ref{fig:j0ce_usage}), because several of the diseases have lags of approximately one year, i.e. 52 weeks. In the top 10 for \textit{Disease names pro} are two chronic diseases: type-2 diabetes and COPD. For both of these the lag does not correspond to a yearly seasonality. This likely indicates that, it is not these precise diseases that are treated by antimicrobials; rather, patients of these diseases are likely to develop weaker immune systems, indicating that after one or two years they are more prone to complications needing antimicrobial treatment.

Similarly to Table \ref{tabel:top10_4weeks}, we also observe in Table \ref{tabel:top10_130weeks}, that \textit{Descriptions lay} and \textit{Description lay frequent} yield words unrelated to antimicrobials as queries. Inspecting the lag of the unrelated words, e.g. 94 weeks for ``growth'', we observe that they are spurious correlations, but not seasonal, as has been previously observed for the prediction of influenza like illnesses \cite{lazer2014parable, dalum2017seasonal}. Such spurious correlations cannot be expected to reliably model rapid changes in antimicrobial drug consumption, as discussed above.

Figures \ref{fig:query_set_performance_web-gp} \& \ref{fig:query_set_performance_web-en} show the prediction error when only using web search data from the different query sets and for different lags, for Gaussian Processes and Elastic Net, respectively. While we previously saw in Tables \ref{tabel:top10_4weeks} \& \ref{tabel:top10_130weeks}, that \textit{Drug names pro} and \textit{Drug names lay} were the data sources with the most semantically relevant queries, we now see that \textit{Descriptions lay} generally is the best performing query set.
We previously observed that \textit{Descriptions lay} contained spurious
correlations, so it seems strange that this query set performs best.
Similar observations have been made with respect to ILI prediction,
where the semantically relevant query set was outperformed by
a less relevant one \cite{dalum2017seasonal}. This likely happens
for two reasons: (i) spurious correlations can model the expected
seasonality well, (ii) lack of evaluation data can make correlations
due to chance more likely \cite{LiomaH17}.

We also see in Figures \ref{fig:query_set_performance_web-gp} \& \ref{fig:query_set_performance_web-en} that there is a noticeable reduction in prediction error, for all query sets, when moving from a maximum lag of 26 weeks to 130 weeks. This is likely due to the modeling of seasonal variations that we noticed in Table \ref{tabel:top10_130weeks}. Even when predicting 12 weeks into the future, the prediction error is still relatively stable. As we saw in Table \ref{tabel:top10_130weeks}, there are long term effects of antimicrobial drug usage, either seasonal changes or long term predictors such as type-2 diabetes, and these are likely some of the reasons why prediction into the future works well.

The impact of query selection upon prediction performance significantly diminishes when prediction is based on a combination of web data and historical antimicrobial purchase data, i.e. when high quality time series data is available. We see that in Figures \ref{fig:query_set_performance_antibiotis-gp} \& \ref{fig:query_set_performance_antibiotis-en}.

Finally, as we noted previously, the consistent prediction performance across a prediction offset of 0 weeks and 12 weeks strong. Figure 1 shows that the three last years of our antimicrobial consumption time series are very similar. This is likely one of the reasons for the consistent performance independent of the prediction offset. It is not unlikely that the prediction models will perform significantly worse in case of a sudden change, as was observed with Google Flu during the 2009 swine flu \cite{cook2011assessing}. In such a scenario we would expect the semantically relevant queries to remain correlated with the consumption, while the search pattern for the irrelevant queries should remain unchanged given changes in antimicrobial consumption. Given such a change in consumption, it is likely that the difference between the \textit{Disease names pro} query set and \textit{Descriptions lay} query set would become apparent.

%
%
%
%

\section{Conclusion}
We studied the extent to which consumption of antimicrobial drugs, such as antibiotics, can be predicted from web search data. We compared this to predictions based on more traditional historical purchase data of antimicrobial drugs. We experimented with different prediction models (Elastic Net and Gaussian Processes), and a novel method of selecting web search queries indicative of antimicrobial drug consumption by mining antimicrobial related information from publicly available descriptions of diseases and drugs linked to antimicrobials. Experiments with more than 9 years of weekly antimicrobial drug consumption data from Denmark showed that prediction using web search data are overall comparable and marginally more erroneous than predictions using antimicrobial drug sales data. The difference in error between the two is equivalent to 1\% point mean absolute error in weekly consumption. This performance was found to be relatively stable across variations in prediction offsets, prediction models, and query selection methods.



\paragraph{\textbf{Competing interests}} The authors have declared that no competing interests exist.

\bibliographystyle{ACM-Reference-Format}
\bibliography{bibliography}

\end{document}